\documentstyle[12pt,subeqnarray,psfig]{article}
\begin{document}
%
%
%
%
\newenvironment{lefteqnarray}{\arraycolsep=0pt\begin{eqnarray}}
{\end{eqnarray}\protect\aftergroup\ignorespaces}
\newenvironment{lefteqnarray*}{\arraycolsep=0pt\begin{eqnarray*}}
{\end{eqnarray*}\protect\aftergroup\ignorespaces}
\newenvironment{leftsubeqnarray}{\arraycolsep=0pt\begin{subeqnarray}}
{\end{subeqnarray}\protect\aftergroup\ignorespaces}
\newcommand{\diff}{{\rm\,d}}
\newcommand{\appleq}{\stackrel{<}{\sim}}
\newcommand{\appgeq}{\stackrel{>}{\sim}}
\newcommand{\Int}{\mathop{\rm Int}\nolimits}
\newcommand{\Nint}{\mathop{\rm Nint}\nolimits}
\newcommand{\range}{{\rm -}}
\newcommand{\displayfrac}[2]{\frac{\displaystyle #1}{\displaystyle #2}}
%
%
\title{The G-dwarf problem in the Galactic spheroid}
\author{{R.~Caimmi}\footnote{
{\it Astronomy Department, Padua Univ., Vicolo Osservatorio 2,
I-35122 Padova, Italy}
email: caimmi@pd.astro.it~~~
fax: 39-049-8278212}
\phantom{agga}}
%
%
\maketitle
\begin{quotation}
\section*{}
\begin{Large}
\begin{center}

Abstract

\end{center}
\end{Large}
\begin{small}

\noindent\noindent
Using two alternative [O/H]-[Fe/H] dependences, the
empirical oxygen abundance distribution (EGD) is 
deduced from two different samples involving (i)
268 K-giant bulge stars (Sadler et al. 1996), and 
(ii) 149 globular clusters (Mackey \& van den Bergh
2005) for which the iron abundance distribution is
known, in addition to previous results (Caimmi 2001)
related to (iii) 372 solar neighbourhood halo
subdwarfs (Ryan \& Norris 1991).   Under the
assumption that each distribution is typical for
the corresponding subsystem, the EGD of the Galactic
spheroid is determined weighting by mass.   The
trend is non monotonic, with the occurrence of
one minimum and two maxima within the domain.
The data are fitted, to an acceptable extent, by
simple models of chemical evolution implying both
homogeneous and inhomogeneous mixing, provided
star formation is inhibited during halo formation
and enhanced during bulge formation, with respect
to the disk.
The theoretical oxygen abundance distribution
(TGD) is first determined for the halo and the
bulge separately, and then for the Galactic
spheroid weighting by mass.   Though a G-dwarf
problem seems to exist for both the halo and
the bulge, it could be alleviated or removed
allowing an increasing star formation efficiency
during the formation of a Galactic subsystem,
which does not necessarily imply gas infall.
The results are independent of the power-law
initial mass function (IMF) exponent, provided
both the lower stellar mass limit and the mass
fraction in long-lived stars and remnants per
star generation, are suitably changed.   Then
the effect of star formation inhibiting or
enhancing gas is counter-balanced.   Simple
models implying homogeneous mixing are unable
in fitting the empirical age-metallicity
relation (EAMR) recently determined from
a homogeneous sample of globular clusters
(De Angeli et al. 2005), which shows a non
monotonic trend characterized by large
dispersion.   On the other hand, simple
models involving inhomogeneous mixing yield
a theoretical age-metallicity relation (TAMR)
which reproduces the data to an acceptable
extent.   With regard to gas ouflow from the
proto-halo, acceptable models make different
predictions according if the Galactic spheroid
and disk underwent separate or common chemical
evolution.   In the former alternative, less
than one third of the bulge mass outflowed
from the proto-halo.   In the latter
alternative, the existence of an unseen
baryonic halo (or equivalent amount of gas
lost by the Galaxy) with mass comparable to
bulge mass, is necessarily needed.   In this
view, the outflowing proto-halo gas which
remains bound to the Galaxy, makes both the
bulge and the disk.   The treshold star mass
below which the halo is not visible (or an
equivalent amount of gas has been lost) is
calculated as $m_0\approx0.25~{\rm m}_\odot$ for
IMF exponent $p=2.9$ and $m_0\approx0.10~
{\rm m}_\odot$ for $p=2.35$.   Conversely, $p
\approx2.8$ for lower limit stellar mass,
$m_{mf}=0.25~{\rm m}_\odot$ and $p\approx2.6$
for $m_0\approx0.10~{\rm m}_\odot$.

\noindent
{\it keywords - 
galaxies: evolution - stars: formation; evolution.}
\end{small}
\end{quotation}

\section{Introduction} \label{intro}

The existence of a G-dwarf problem i.e. the observation
of too few metal deficient G dwarfs (or, more generally,
of a selected spectral type) with respect to what expected
from the Simple model of chemical evolution (e.g., Searle
\& Sargent 1972; Pagel \& Patchett 1975; Haywood 2001) was
first established in the solar neighbourhood (van den Bergh
1962; Schmidt 1963).   Though in less extreme form, a
G-dwarf problem appears to exist in both the halo (e.g.,
Hartwick 1976; Prantzos 2003) and in the bulge (e.g.,
Ferreras, Wyse \&
Silk 2003).   In addition, a G-dwarf problem has been
detected in both bulge-dominated and disk-dominated
galaxies (Henry \& Worthey 1999), which is consistent
with the idea that the G-dwarf problem is universal
(Worthey, Dorman \& Jones 1996).

The deficit of metal-poor stars (with respect to the
prediction of the Simple model) may be interpreted
in different ways, such as changes in the initial
mass function (Schmidt 1963; Adams \& Fatuzzo 1996;
Bromm 2004; Bromm \& Larson 2004; Larson 2005),
inflow of unprocessed (Larson 1974) or processed
(Thacker, Scannapieco \& Davis 2002) material from
outside, or evolution with inhomogeneous mixing
(Searle 1972; Malinie et al. 1993; Caimmi 2000,
2001b, hereafter quoted as C00%
\footnote{With regard to this reference, two points
may carefully be kept in mind, namely (i) values of
a few parameters must be corrected as explained in
Caimmi (2001b), Sect.\,3, second paragraph therein,
and (ii) the majority of figures do not correspond
to the caption, as explained in the erratum (Caimmi
2001a).}
 and C01, respectively; Oey 2003; Karlsson 2005).
For additional alternatives and further details
see e.g., Pagel \& Patchett (1975); Pagel (1989).

In addition to the G-dwarf problem, a lack of a
well-defined empirical age-metallicity relation
(EAMR) seems to be established for both the disk
solar neighbourhood (e.g., Meusinger, Reimann \&
Stecklum 1991; Edvardsson et al. 1993; Rocha-Pinto
et al. 2000; Feltzing, Holmberg \& Hurley 2001;
Nordstr\"om et al. 2004; Karatas, Bilitz \& Schuster
2005) and globular clusters (Salaris \& Weiss 2002;
De Angeli et al. 2005).   The large scatter observed
in the EAMR is probably universal, at least with
regard to massive enough ($M\appgeq10^{10}{\rm m}_\odot$)
galaxies, independent of the morphological type.

Inhomogeneous (i.e. implying inhomogeneous gas
mixing) models of chemical evolution succeed in
both providing a solution to the G-dwarf problem
and reproducing substantial scatter exhibited by
the EAMR.   The current paper aims to investigate
if inhomogeneous simple models of chemical
evolution are also consistent with the metallicity
distribution in the Galactic spheroid, deduced
weighting by mass data belonging to subsystems,
more specifically solar neighbourhood halo
subdwarfs, K-giant bulge stars, and globular
clusters.   In this view, the next step is to
see what constraints are related to the formation
and the evolution of the Galaxy.

With regard to halo subdwarfs, the oxygen abundance
has been deduced (C01) from data related to a sample
of 372 kinematically selected halo stars (Ryan \&
Norris 1991) after conversion of [Fe/H] into [O/H]
using two alternative empirical relations, involving
the presence or absence of [O/Fe] plateau for
sufficiently low [Fe/H] values.

With regard to globular clusters and bulge stars, a 
similar procedure is applied to: (i) a homogeneous 
sample of 55 objects for which the EAMR is also 
known (De Angeli et al. 2005), (ii) an inhomogeneous 
sample of 149 objects (Mackey \& van den Bergh 2005),
and (iii) a homogeneous sample of 268 bulge K-giant
(Sadler, Rich \& Terndrup 1996), as reported in 
Section 2.   In addition, it is derived therein a
putative oxygen abundance distribution in all halo
objects, under the assumption that both globular clusters
and field stars underwent a common chemical evolution.
The same is made also for the Galactic spheroid, with
the inclusion of bulge stars.   A comparison with the
predictions of simple models, involving homogeneous and
inhomogeneous mixing, is made in Section 3 and 4, 
respectively.   The discussion and the conclusion
are the subject of Section 5 and 6, respectively.

\section{The data} \label{data}
\subsection{Empirical age-metallicity relation in
globular clusters} \label{EAMR}

Accurate relative ages for a sample of 55 Galactic
globular clusters have recently been determined 
(De Angeli et al. 2005).   The ages were obtained
by measuring the difference between the horizontal
branch and the turnoff in two internally 
photometrically homogeneous databases with 16
objects in common.   The related EAMR is derived
from absolute ages and related errors (De Angeli
2005) for objects belonging to a single database.
With regard to objects in common where a value 
and an error exist for each database, the absolute
age is calculated as the centre of the intersection
of the two intervals, and the error as the 
corresponding semiamplitude.   For instance, 
$11.42\mp0.34$ and $10.96\mp0.60$ yield $11.32\mp
0.24$%
\footnote{The absolute age could be calculated as a 
weighted mean, provided the errors available for
each database may be related to empirical variances.}.
Following De Angeli et al. (2005), metallicities
are calibrated over two different scales, namely
CG (Carretta \& Gratton 1997, as extended by
Carretta et al. 2001) and ZW (Zinn \& West 1984).
In addition to the above homogeneous values, a
third alternative consists in using the data
(taken from different sources) from Harris (1996)
catalogue (2003 update), also listed (with one
exception) in a recent paper (Mackey \& van den
Bergh 2005), where five classification types
are defined.

With regard to the current sample of 55 objects,
three classification types among the above
mentioned five shall be retained, namely:
BD for bulge/disk clusters (5 objects); OH
for old halo clusters (36 objects); and YH 
for young halo clusters (13 objects).   For
further details, see Mackey \& van den Bergh
(2005).   An additional cluster (NGC 6366)
has to be considered by itself, as belonging
to BD type ([Fe/H]$>-$0.8) with respect to CG 
and ZW metallicity scales, and to OH type
([Fe/H]$\le-$0.8) with resepct to the value
listed in Harris catalogue (Mackey \& van
den Bergh 2005).

\begin{figure}[t]
\centerline{\psfig{file=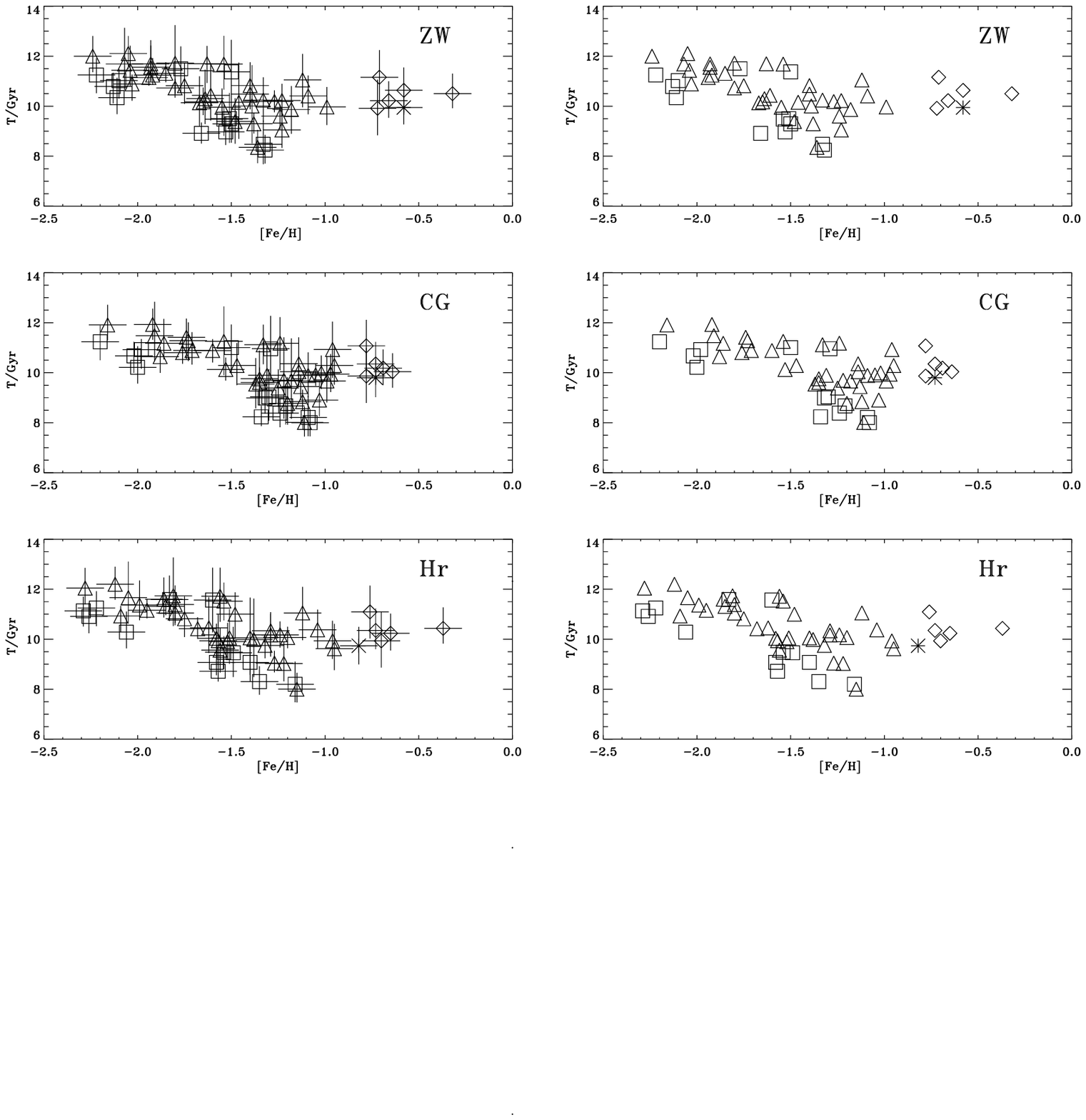,height=130mm,width=140mm}}
\caption[mamr]{Empirical age-metallicity relation
(EAMR) from a sample of 55 globular clusters (De
Angeli et al. 2005) in terms of absolute ages (De
Angeli 2005), with regard to ZW (top panels) and 
CG (middle panels) metallicity scales, and values 
from Harris catalogue (Mackey \& van den Bergh 2005;
bottom panels), with (left panels) and without
(right panels) error bars.   Morphological types:
OH (old halo) - triangles; YH (young halo) - 
squares; BD (bulge/disk) - diamonds; NCG6366 -
asterisk.}
\label{f:MAMR}
\end{figure}

The EAMR for the sample under discussion is
represented in Fig.\,\ref{f:MAMR}, with regard 
to the above mentioned metallicity scales.
An inspection to Fig.\,\ref{f:MAMR} discloses the
following.
\begin{description}
\item[\rm{(i)}]
Age differences related to different metallicity
scales are small.
\item[\rm{(ii)}]
The formation of globular clusters in the
Galactic halo was a continuous process which
span over $\approx4~$Gyr.
Low-metallicity halo clusters were generated
over $\approx2~$Gyr.   Intermediate-metallicity
halo clusters were generated over $\approx4~$Gyr.
High-metallicity halo clusters and bulge/disk
clusters were generated over $\approx1~$Gyr.
\item[\rm{(iii)}]
No clear distinction between ``old halo''
and ``young halo'' clusters seems to be in
terms of age difference.   Perhaps it should
be better referring to ``primeval halo'' and
``accreted halo'' clusters.   The inclusion
of YH clusters has no appreciable effect on
the EAMR.
\end{description}

Accordingly, the following picture of Galactic
evolution may be inferred.   Low-metallicity
stars formed within the proto-halo and after
about 1 Gyr within the proto-bulge/proto-disk.
Bulge formation lasted about 1 Gyr, consistent
with the absence of bulge stars younger than
10 Gyr (Zoccali et al. 2003), while halo
formation ended about 8 Gyr ago, similar
to what has been found for the thick disk
(Liu \& Chaboyer 2000; Ibukiyaua \& Arimoto
2002).   Then halo and thick disk formation
lasted about 4 Gyr and 3 Gyr, respectively.
In addition, one has to keep in mind that
different specific angular momentum distributions
occur in halo-bulge and thick disk-thin disk
stars, which implies that the two subsystems
had different dynamical evolutions (Ibata \&
Gilmore 1995).

\subsection{Metallicity distribution in
globular clusters and field stars}\label{medi}

The empirical distribution of oxygen abundance in
globular clusters is deduced
from a sample of 149 globular clusters where
[Fe/H] has been determined from different sources
(Mackey \& van den Bergh 2005) and from a sample of
55 globular clusters where [Fe/H] has been determined
from a single source (De Angeli et al. 2005). 
In dealing with
simple models of chemical evolution, involving the
assumption of instantaneous recycling, the predicted
metal abundance has to be compared with the observed
oxygen abundance (e.g., Pagel 1989; C00; C01).
Unfortunately, oxygen is more difficult than
iron to detect, and an empirical relation is needed,
to express the former as a function of the latter.

To this respect, a clear
dichotomy appears to be among authors who
support a plateau in [O/Fe] for stars with
[Fe/H]$\appleq-1$ (e.g., Carretta, Gratton \& Sneden
2000); and those who do not (e.g., Israelian
et al. 2001); to get further insight, see
the proceedings edited by Barbuy et al.
(2001).   For exploiting both the above 
possibilities into consideration, the 
following relations (C01) shall be used:
\begin{lefteqnarray}
\label{eq:gra}
&& \left[\frac{\rm O}{\rm H}\right]=
\cases{\left[\displayfrac{\rm Fe}{\rm H}\right]+0.6~;
& $\left[\displayfrac{\rm Fe}{\rm H}\right]\le-1.2$ ; 
$\quad\left[\displayfrac{\rm O}{\rm H}\right]\le-0.6$ ; \cr
& \cr
\displayfrac12\left[\displayfrac{\rm Fe}{\rm H}\right]~;
& $\left[\displayfrac{\rm Fe}{\rm H}\right]\ge-1.2$ ; 
$\quad\left[\displayfrac{\rm O}{\rm H}\right]\ge-0.6$ ;\cr} \\
&& \nonumber \\
\label{eq:isa}
&& \left[\frac{\rm O}{\rm H}\right]=\frac23\left[\frac{\rm Fe}{\rm H}\right]~;
\end{lefteqnarray}
in presence or in absence of [Fe/H] plateau,
respectively, where the oxygen solar abundance
is taken to be ${\rm O}_\odot=0.0056$ (Allende-Prieto,
Lambert, \& Asplund 2001); for further details, see C01.

A number of [O/H]-[Fe/H] relations lying
between those
expressed by Eqs.\,(\ref{eq:gra}) and 
(\ref{eq:isa}), have been derived from
recent investigations (Jonsell et al.
2005; Fulbright, Rich \& McWilliam 2005;
Garcia Perez et al. 2005; Melendez et al.
2005).
To allow comparison with previous
results related to halo subdwarfs (C01),
Eqs.\,(\ref{eq:gra}) and (\ref{eq:isa})
shall be used in the current attempt,
and hereafter quoted as ``in presence''
and ``in absence'' of [O/Fe] plateau,
respectively, with regard to sufficiently 
low metallicities, [Fe/H]$\appleq-1$.

On the other hand, extremely metal deficient
([Fe/H]$<-4$) stars are known to be oxygen
overabundant (e.g., Christlieb et al. 2002;
Iwamoto et al. 2005), [O/Fe]$\approx
2.5$ (Bromm \& Loeb 2003; Frabel et al.
2006).   Then Eqs.\,(\ref{eq:gra}) and
(\ref{eq:isa}) cannot be used in this case,
and the chemical evolution shall be restricted
to a later epoch where only pop.\,II star
formation occurred, [Fe/H]$>-4$, say.

Let us define the oxygen abundance
normalized to the solar value, $\phi$, as:
\begin{equation}
\label{eq:lgfi}
\log\phi=\log\frac{\rm O}{{\rm O}_\odot}=
\left[\frac{\rm O}{\rm H}\right]~~;
\end{equation}
and let $\Delta\log\phi=\Delta$[O/H]=[O/H]$^+-$
[O/H]$^-$
be a logarithmic, oxygen abundance bin deduced
from $\Delta$[Fe/H] by use of Eq.\,(\ref{eq:gra})
or (\ref{eq:isa}).   The related, oxygen 
abundance bin is:
\begin{leftsubeqnarray}
\slabel{eq:fiba}
&& \Delta\phi=\Delta^+\phi+\Delta^-\phi~;\quad\Delta^
\mp\phi=\vert\phi-\phi^\mp\vert~~; \\
\slabel{eq:fibb}
&& \phi^\mp=\exp_{10}\left[\frac{\rm O}{\rm H}\right]^\mp~;\quad
\phi=\frac{\phi^++\phi^-}2~~;
\label{seq:fib}
\end{leftsubeqnarray}
where in general, $\exp_\xi$ defines the power
of basis $\xi$ and, in particular, $\exp$
defines the power of basis e, according to the
standard notation.   As in C01, bins in [Fe/H]
equal to 0.2 dex shall be used (e.g., Norris
\& Ryan 1991; Huchra, Brodie \& Kent 1991;
Perrett et al. 2002).

The empirical, differential metallicity distribution
(hereafter referred to as EGD) in a selected class
of objects, is defined as (Pagel 1989; C00; C01):
\begin{equation}
\label{eq:psi}
\psi(\phi\mp\Delta^\mp\phi)=\log\frac{\Delta N}{N\Delta
\phi} ~;
\end{equation}
where $\Delta\phi$ is the bin width,
$\Delta N$ is the number of sample objects with
oxygen abundance belonging to a bin centered in
$\phi$, and $N$ is the total number of sample
objects.   The differential distribution is
used instead of the cumulative distribution, 
as it is a more sensitive test (Pagel 1989)
and allows direct comparison between different
samples.   The uncertainty on $\Delta N$ 
has been evaluated from Poisson errors (e.g., 
Ryan \& Norris 1991), as $\Delta(\Delta N)=
(\Delta N)^{1/2}$, and the related uncertainty 
in the EGD is (e.g., C01):
\begin{leftsubeqnarray}
\slabel{eq:psiera}
&& \Delta^\mp\psi=\vert\psi-\psi^\mp\vert=
\left\vert\log\left[1\mp\frac{(\Delta N)^
{1/2}}{\Delta N}\right]\right\vert~~; \\
\slabel{eq:psierb}
&& \psi^\mp=\log\frac{\Delta N\mp(\Delta N)^{1/2}}{N\Delta
\phi}~~;
\label{seq:psier}
\end{leftsubeqnarray}
where $\psi^-$ diverges to $-\infty$ in the limit
$\Delta N\rightarrow1$.   For further details,
see C01.

The [Fe/H]-[O/H] relation and corresponding
mean fractional oxygen abundance, $\phi$,
and half bin width, $\Delta^\mp\phi$, in
presence of [O/Fe] plateau (PP), according
to Eq.\,(\ref{eq:gra}), and in absence of
[O/Fe] plateau (AP), according to 
Eq.\,(\ref{eq:isa}), respectively,
are shown in Tab.\,\ref{t:OFe} for
the metallicity range of interest.
\begin{table}
\caption[par]{The [Fe/H]-[O/H] relation
and corresponding
mean fractional oxygen abundance, $\phi$,
and half bin width, $\Delta^\mp\phi$, in
presence (PP) and in absence (AP) of [O/Fe] 
plateau.   To save space, $F$ stays for
[Fe/H] and $O$ for 3[O/H].}
\label{t:OFe}
\begin{center}
\begin{tabular}{rrrrrrrrrr}
\multicolumn{2}{c|}{}
& \multicolumn{4}{c|}{PP}
& \multicolumn{4}{c}{AP} \\
\hline\noalign{\smallskip}
\multicolumn{1}{c}{$\phantom{88}F^-$} &
\multicolumn{1}{c}{\phantom{88}$F^+$} &
\multicolumn{1}{c}{\phantom{88}$O^-$} &
\multicolumn{1}{c}{\phantom{88}$O^+$} &
\multicolumn{1}{c}{$\phi$} & \multicolumn{1}{c}{$\Delta^\mp\phi$} &
\multicolumn{1}{c}{\phantom{88}$O^-$} &
\multicolumn{1}{c}{\phantom{88}$O^+$} &
\multicolumn{1}{c}{$\phi$} & \multicolumn{1}{c}{$\Delta^\mp\phi$} \\
\noalign{\smallskip}
\hline\noalign{\smallskip}
1.2 & 1.4 & 1.8 & 2.1 & 4.496 & 0.515 & 2.4 & 2.8 & 7.443 & 1.134 \\
1.0 & 1.2 & 1.5 & 1.8 & 3.572 & 0.409 & 2.0 & 2.4 & 5.476 & 0.834 \\
0.8 & 1.0 & 1.2 & 1.5 & 2.837 & 0.325 & 1.6 & 2.0 & 4.028 & 0.613 \\
0.6 & 0.8 & 0.9 & 1.2 & 2.254 & 0.258 & 1.2 & 1.6 & 2.963 & 0.451 \\
0.4 & 0.6 & 0.6 & 0.9 & 1.790 & 0.205 & 0.8 & 1.2 & 2.180 & 0.332 \\
0.2 & 0.4 & 0.3 & 0.6 & 1.422 & 0.163 & 0.4 & 0.8 & 1.604 & 0.244 \\
0.0 & 0.2 & 0.0 & 0.3 & 1.129 & 0.129 & 0.0 & 0.4 & 1.180 & 0.180 \\
$-$0.2 & 0.0 & $-$0.30 & 0.0 & 0.897 & 0.103 & $-$0.4 & 0.0 & 0.868 
& 0.132 \\
$-$0.4 & $-$0.2 & $-$0.6 & $-$0.3 & 0.713 & 0.087 & $-$0.8 & $-$0.4 & 
0.638 & 0.097 \\
$-$0.6 & $-$0.4 & $-$0.9 & $-$0.6 & 0.566 & 0.065 & $-$1.2 & $-$0.8 & 
0.470 & 0.071 \\
$-$0.8 & $-$0.6 & $-$1.2 & $-$0.9 & 0.450 & 0.052 & $-$1.6 & $-$1.2 & 
0.341 & 0.053 \\
$-$1.0 & $-$0.8 & $-$1.5 & $-$1.2 & 0.357 & 0.041 & $-$2.0 & $-$1.6 & 
0.254 & 0.039 \\
$-$1.2 & $-$1.0 & $-$1.8 & $-$1.5 & 0.284 & 0.032 & $-$2.4 & $-$2.0 & 
0.187 & 0.028 \\
$-$1.4 & $-$1.2 & $-$2.4 & $-$1.8 & 0.205 & 0.046 & $-$2.8 & $-$2.4 & 
0.137 & 0.021 \\
$-$1.6 & $-$1.4 & $-$3.0 & $-$2.4 & 0.129 & 0.029 & $-$3.2 & $-$2.8 & 
0.101 & 0.015 \\
$-$1.8 & $-$1.6 & $-$3.6 & $-$3.0 & 0.081 & 0.018 & $-$3.6 & $-$3.2 & 
0.074 & 0.011 \\
$-$2.0 & $-$1.8 & $-$4.2 & $-$3.6 & 0.051 & 0.012 & $-$4.0 & $-$3.6 & 
0.055 & 0.008 \\
$-$2.2 & $-$2.0 & $-$4.8 & $-$4.2 & 0.032 & 0.007 & $-$4.4 & $-$4.0 & 
0.040 & 0.006 \\
$-$2.4 & $-$2.2 & $-$5.4 & $-$4.8 & 0.020 & 0.005 & $-$4.8 & $-$4.4 & 
0.030 & 0.004 \\
$-$2.6 & $-$2.4 & $-$6.0 & $-$5.4 & 0.013 & 0.003 & $-$5.2 & $-$4.8 & 
0.022 & 0.003 \\
$-$2.8 & $-$2.6 & $-$6.6 & $-$6.0 & 0.008 & 0.002 & $-$5.6 & $-$5.2 & 
0.016 & 0.002 \\
$-$3.0 & $-$2.8 & $-$7.2 & $-$6.6 & 0.005 & 0.002 & $-$6.0 & $-$5.6 & 
0.012 & 0.002 \\
$-$3.7 & $-$3.0 & $-$9.3 & $-$7.2 & 0.002 & 0.002 & $-$7.4 & $-$6.0 & 
0.007 & 0.003 \\
\noalign{\smallskip}
\hline
\end{tabular}
\end{center}
\end{table}
The EGD derived from the sample studied by
De Angeli et al. (2005), using Eqs.\,(\ref
{eq:lgfi}), (\ref{seq:fib}), (\ref{eq:psi}),
(\ref{seq:psier}), is listed in Tab.\,\ref
{t:E55} in presence of [O/Fe] plateau with
regard to CG metallicity calibration, and
in absence of [O/Fe] plateau with
regard to ZW metallicity calibration.
\begin{table}
\caption[par]{The empirical, differential
metallicity distribution (EGD) in globular 
clusters, deduced from a sample of
55 objects studied by De Angeli et al.
(2005), in presence of [O/Fe] plateau
(PP) with regard to CG metallicity 
calibration, and in absence of [O/Fe] 
plateau (AP) with regard to ZW metallicity 
calibration.}
\label{t:E55}
\begin{center}
\begin{tabular}{rrrrrrrrrr}
\multicolumn{5}{c|}{PP}
&\multicolumn{5}{c}{AP} \\
\hline\noalign{\smallskip}
\multicolumn{1}{c}{$\phi$} & \multicolumn{1}{c}{$\phantom{0}\psi$} &
\multicolumn{1}{c}{$\Delta^-\psi$} & \multicolumn{1}{c}{$\Delta^+\psi$} &
\multicolumn{1}{c}{$\Delta N$} & \multicolumn{1}{c}{$\phi$} &
\multicolumn{1}{c}{$\phantom{0}\psi$} &
\multicolumn{1}{c}{$\Delta^-\psi$} & \multicolumn{1}{c}{$\Delta^+\psi$} &
\multicolumn{1}{c}{$\Delta N$}  \\
\noalign{\smallskip}
\hline\noalign{\smallskip}
0.713 & 0.000 & 0.000 & 0.000 & \phantom{0}0 &
0.638 & $-$1.029 & $\infty$\phantom{0.} & 0.301 & \phantom{0}1 \\
0.566 & 0.000 & 0.000 & 0.000 & 0 &
0.470 & $-$0.896 & $\infty$\phantom{0.} & 0.301 & \phantom{0}1 \\
0.450 & 0.025 & 0.228 & 0.149 & 6 &
0.341 & $-$0.160 & 0.301 & 0.176 & \phantom{0}4 \\
0.357 & $-$0.051 & 0.301 & 0.176 & 4 &
0.254 & $-$0.629 & $\infty$\phantom{0.} & 0.301 & \phantom{0}1 \\
0.284 & 0.526 & 0.148 & 0.110 & 12 &
0.187 & $-$0.019 & 0.374 & 0.198 & \phantom{0}3 \\
0.205 & 0.469 & 0.130 & 0.100 & 15 & 
0.137 & 0.592 & 0.176 & 0.125 & \phantom{0}9 \\
0.129 & 0.095 & 0.301 & 0.176 & \phantom{0}4 &
0.101 & 0.885 & 0.141 & 0.106 & \phantom{}13 \\
0.081 & 0.391 & 0.257 & 0.160 & \phantom{0}5 &
0.074 & 0.807 & 0.189 & 0.131 & \phantom{0}8 \\
0.051 & 0.591 & 0.257 & 0.160 & \phantom{0}5 &
0.055 & 0.816 & 0.228 & 0.149 & \phantom{0}6 \\
0.032 & 0.570 & 0.374 & 0.198 & \phantom{0}3 &
0.040 & 1.016 & 0.206 & 0.139 & \phantom{0}7 \\
0.020 & 0.293 & $\infty$\phantom{0.} & 0.301 & \phantom{0}1 &
0.030 & 0.605 & 0.533 & 0.232 & \phantom{0}2 \\
\noalign{\smallskip}
\hline
\end{tabular}
\end{center}
\end{table}

\begin{figure}[t]
\centerline{\psfig{file=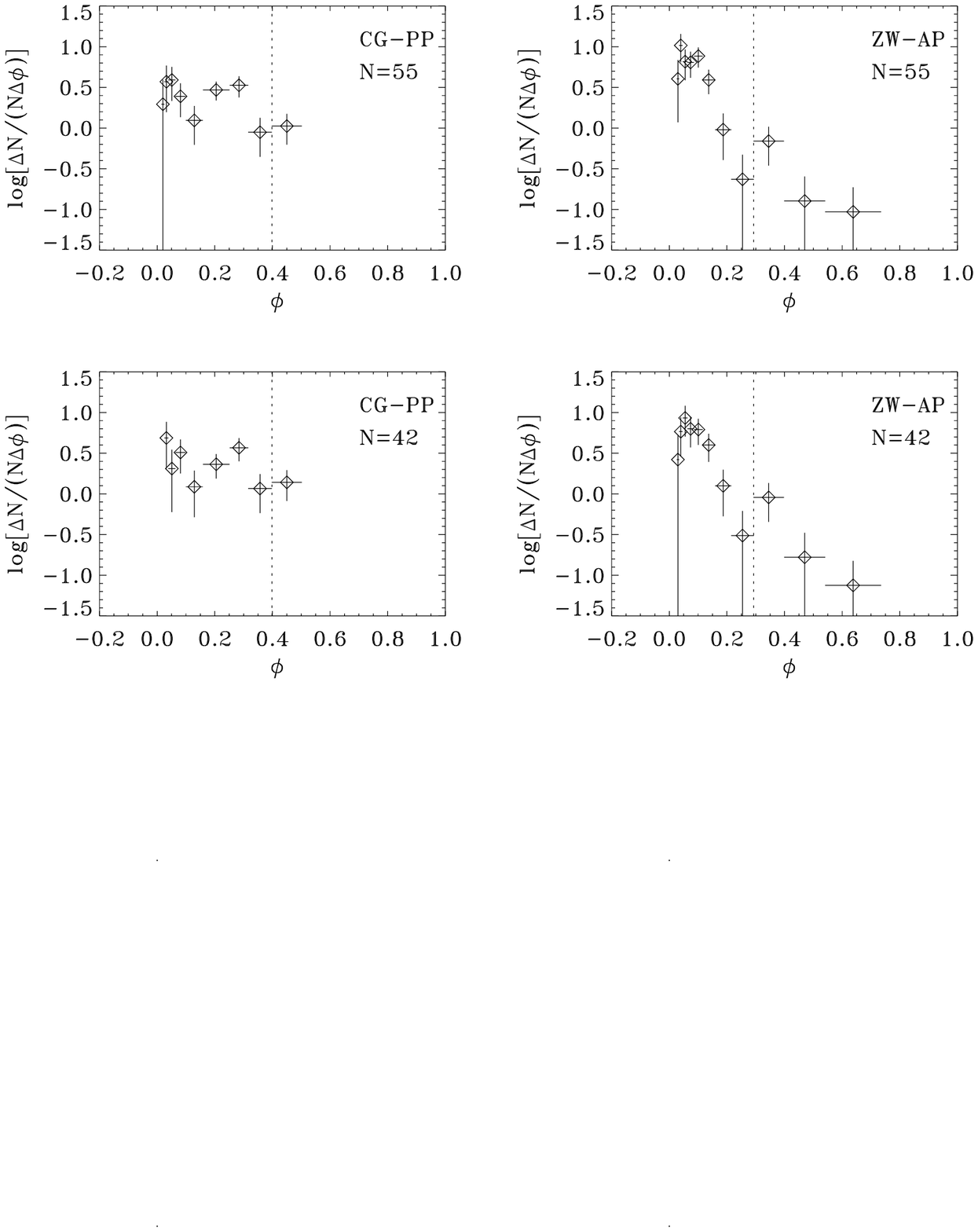,height=130mm,width=140mm}}
\caption[EGD]{The empirical, differential metallicity
distribution (EGD) in globular clusters, plotted with
regard to a complete sample (De Angeli et al. 2005;
$N=55$, top panels) and a reduced sample ($N=42$,
bottom panels) with YH clusters removed, both in 
presence (left panels, CG metallicity calibration)
and in absence (right panels, ZW metallicity
calibration), of [O/Fe] plateau, respectively.
The dotted vertical line marks the transition from
halo (OH, YH) to bulge/disk (BD) morphological
type, [Fe/H]=$-$0.8.   The distribution appears
to be bimodal, with the occurrence of two maxima,
close to the beginning of evolution and to the
transition from halo to bulge/disk morphological
type, respectively.}
\label{f:EGD}
\end{figure}

The related plots are shown in Fig.\,\ref{f:EGD},
top left and top right, respectively.   The dotted
vertical line marks the transition from halo (OH,
YH) to bulge/disk (BD) morphological type.   Bottom
left and bottom right panels represent a reduced
sample of 42 objects, where YH clusters have been
removed.
The inclusion of YH clusters (safely stripped
from accreted dwarf galaxies like Sagittarius)
appears to have no appreciable effect on the
EGD.   The distribution is bimodal with the
occurrence of two maxima, close to the beginning
of the evolution and to the transition from halo
to bulge/disk morphological type, respectively.

The EGD derived from the sample studied by
Mackey \& van den Bergh (2005), using Eqs.\,(\ref
{eq:lgfi}), (\ref{seq:fib}), (\ref{eq:psi}),
(\ref{seq:psier}), is listed in Tab.\,\ref
{t:E149} in presence and
in absence of [O/Fe] plateau, with
regard to metallicity values taken from
different sources.   
\begin{table}
\caption[par]{The empirical, differential
metallicity distribution (EGD) in globular 
clusters, deduced from a sample of
149 objects studied by Mackey \& van den
Bergh (2005), both in presence (PP) and in 
absence (AP) of [O/Fe] plateau.}
\label{t:E149}
\begin{center}
\begin{tabular}{rrrrrrr}
\multicolumn{2}{c|}{PP}
&\multicolumn{2}{c|}{AP} \\
\hline\noalign{\smallskip}
\multicolumn{1}{c}{$\phi$} & \multicolumn{1}{c}{$\phantom{0}\psi$} &
\multicolumn{1}{c}{$\phi$} & \multicolumn{1}{c}{$\phantom{0}\psi$} &
\multicolumn{1}{c}{$\Delta^-\psi$} & \multicolumn{1}{c}{$\Delta^+\psi$} &
\multicolumn{1}{c}{$\Delta N$}  \\
\noalign{\smallskip}
\hline\noalign{\smallskip}
1.422 & $-$1.686 & 1.604 & $-$1.862 & $\infty$\phantom{0.} & 0.301 &
\phantom{0}1 \\
1.129 & $-$1.785 & 1.180 & $-$1.428 & 0.533 & 0.232 & \phantom{0}2 \\
0.897 & $-$1.185 & 0.868 & $-$1.294 & 0.533 & 0.232 & \phantom{0}2 \\
0.713 & $-$0.784 & 0.638 & $-$1.161 & 0.301 & 0.176 & \phantom{0}4 \\
0.566 & $-$0.140 & 0.470 & $-$0.183 & 0.135 & 0.103 & \phantom{}14 \\
0.450 & $-$0.040 & 0.341 & $-$0.049 & 0.135 & 0.103 & \phantom{}14 \\
0.357 & $-$0.132 & 0.254 & $-$0.108 & 0.176 & 0.125 & \phantom{0}9 \\
0.284 & $-$0.083 & 0.187 & $-$0.026 & 0.189 & 0.131 & \phantom{0}8 \\
0.205 & 0.090 & 0.137 & 0.435 & 0.121 & 0.094 & \phantom{}17 \\
0.129 & 0.491 & 0.101 & 0.769 & 0.093 & 0.076 & \phantom{}27 \\
0.081 & 0.490 & 0.074 & 0.702 & 0.121 & 0.094 & \phantom{}17 \\
0.051 & 0.782 & 0.055 & 0.927 & 0.107 & 0.086 & \phantom{}21 \\
0.032 & 0.563 & 0.040 & 0.641 & 0.189 & 0.131 & \phantom{0}8 \\
0.020 & 0.559 & 0.030 & 0.570 & 0.257 & 0.160 & \phantom{0}5 \\
\noalign{\smallskip}
\hline
\end{tabular}
\end{center}
\end{table}

\begin{figure}[t]
\centerline{\psfig{file=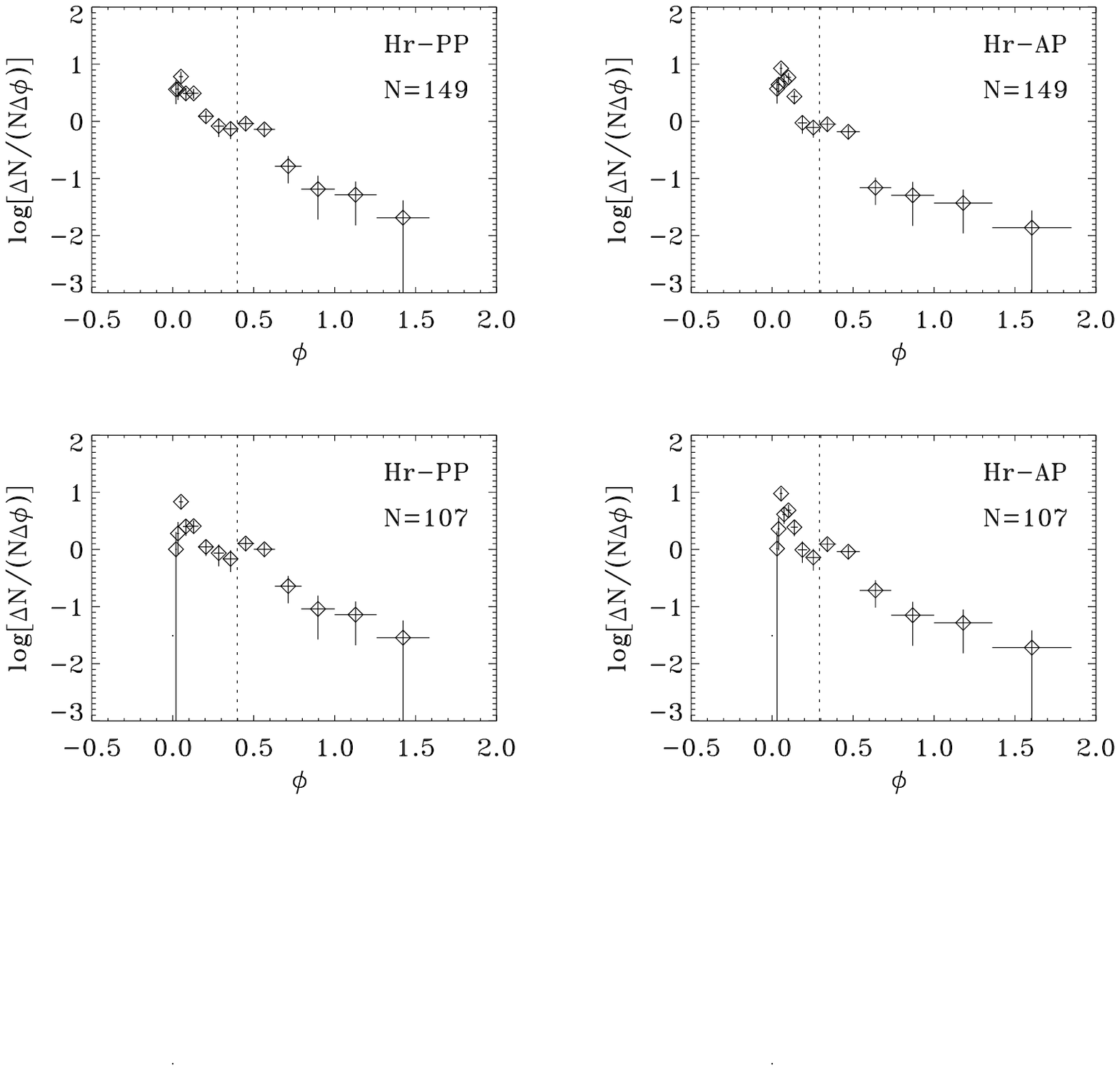,height=130mm,width=140mm}}
\caption[EGD]{The empirical, differential metallicity
distribution (EGD) in globular clusters, plotted with
regard to a complete sample (Mackey \& van den Bergh 2005;
$N=149$, top panels) and a reduced sample ($N=107$,
bottom panels) with only OH clusters retained, both in 
presence (left panels) and in absence (right panels) of 
[O/Fe] plateau, respectively.
The dotted vertical line marks the transition from
halo (OH, YH) to bulge/disk (BD) morphological
type, [Fe/H]$=-$0.8.   The distribution appears
to be bimodal, with the occurrence of two maxima,
close to the beginning of evolution and to the
transition from halo to bulge/disk morphological
type, respectively.}
\label{f:EGD2}
\end{figure}

The related plots are shown in Fig.\,\ref{f:EGD2},
top left and top right, respectively.     Bottom
left and bottom right panels represent a reduced
sample of 107 objects, where only OH clusters have 
been retained.
The inclusion of YH clusters 
appears to have no appreciable effect on the
EGD which, on the other hand, is bimodal with the
occurrence of two maxima, close to the beginning
of the evolution and to the transition from halo
to bulge/disk morphological type, respectively.

The EGD derived from the sample of 268 K-giant
bulge stars in Baade's Window studied by Sadler,
Rich \& Terndrup (1996)%
\footnote{The abundance distribution of the
sample stars has recently been revised, but
with no substantial change (Fulbright et al.
2005, 2006).},
using Eqs.\,(\ref
{eq:lgfi}), (\ref{seq:fib}), (\ref{eq:psi}),
(\ref{seq:psier}), is listed in Tab.\,\ref
{t:B262} both in presence and
in absence of [O/Fe] plateau.
\begin{table}
\caption[par]{The empirical, differential
metallicity distribution (EGD) in K-giant 
bulge stars in the Baade's Window, deduced 
from a sample of 268 objects studied by
Sadler et al. (1996), both in presence
(PP) and in absence (AP) of [O/Fe] 
plateau.}
\label{t:B262}
\begin{center}
\begin{tabular}{rrrrrrr}
\multicolumn{2}{c|}{PP}
&\multicolumn{2}{c|}{AP} \\
\hline\noalign{\smallskip}
\multicolumn{1}{c}{$\phi$} & \multicolumn{1}{c}{$\phantom{0}\psi$} &
\multicolumn{1}{c}{$\phi$} & \multicolumn{1}{c}{$\phantom{0}\psi$} &
\multicolumn{1}{c}{$\Delta^-\psi$} & \multicolumn{1}{c}{$\Delta^+\psi$} &
\multicolumn{1}{c}{$\Delta N$}  \\
\noalign{\smallskip}
\hline\noalign{\smallskip}
4.496 & $-$2.130 & 7.443 & $-$2.473 & 0.533 & 0.232 & \phantom{0}2 \\
3.572 & $-$2.030 & 5.476 & $-$2.339 & 0.533 & 0.232 & \phantom{0}2 \\
2.837 & $-$1.328 & 4.028 & $-$1.604 & 0.189 & 0.131 & \phantom{0}8 \\
2.254 & $-$0.927 & 2.963 & $-$1.170 & 0.125 & 0.097 & \phantom{}16 \\
1.790 & $-$0.569 & 2.180 & $-$0.778 & 0.089 & 0.074 & \phantom{}29 \\
1.422 & $-$0.269 & 1.604 & $-$0.444 & 0.069 & 0.060 & \phantom{}46 \\
1.129 & $-$0.275 & 1.180 & $-$0.417 & 0.079 & 0.067 & \phantom{}36 \\
0.897 & $-$0.175 & 0.868 & $-$0.284 & 0.079 & 0.067 & \phantom{}36 \\
0.713 & $-$0.169 & 0.638 & $-$0.245 & 0.089 & 0.074 & \phantom{}29 \\
0.566 & $-$0.100 & 0.470 & $-$0.142 & 0.093 & 0.076 & \phantom{}27 \\
0.450 & $-$0.201 & 0.341 & $-$0.210 & 0.121 & 0.094 & \phantom{}17 \\
0.357 & $-$0.377 & 0.254 & $-$0.353 & 0.176 & 0.125 & \phantom{0}9 \\
0.284 & $-$0.930 & 0.187 & $-$0.873 & 0.533 & 0.232 & \phantom{0}2 \\
0.205 & $-$1.385 & 0.137 & $-$1.040 & $\infty$\phantom{0.} & 0.301 &
\phantom{0}1 \\
0.129 & $-$1.185 & 0.101 & $-$0.907 & $\infty$\phantom{0.} & 0.301 &
\phantom{0}1 \\
0.081 & 0.000 & 0.074 & 0.000 & 0.000 & 0.000 &  \phantom{0}0 \\
0.051 & $-$0.785 & 0.055 & $-$0.640 & $\infty$\phantom{0.} & 0.301 &
\phantom{0}1 \\
\noalign{\smallskip}
\hline
\end{tabular}
\end{center}
\end{table}

\begin{figure}[t]
\centerline{\psfig{file=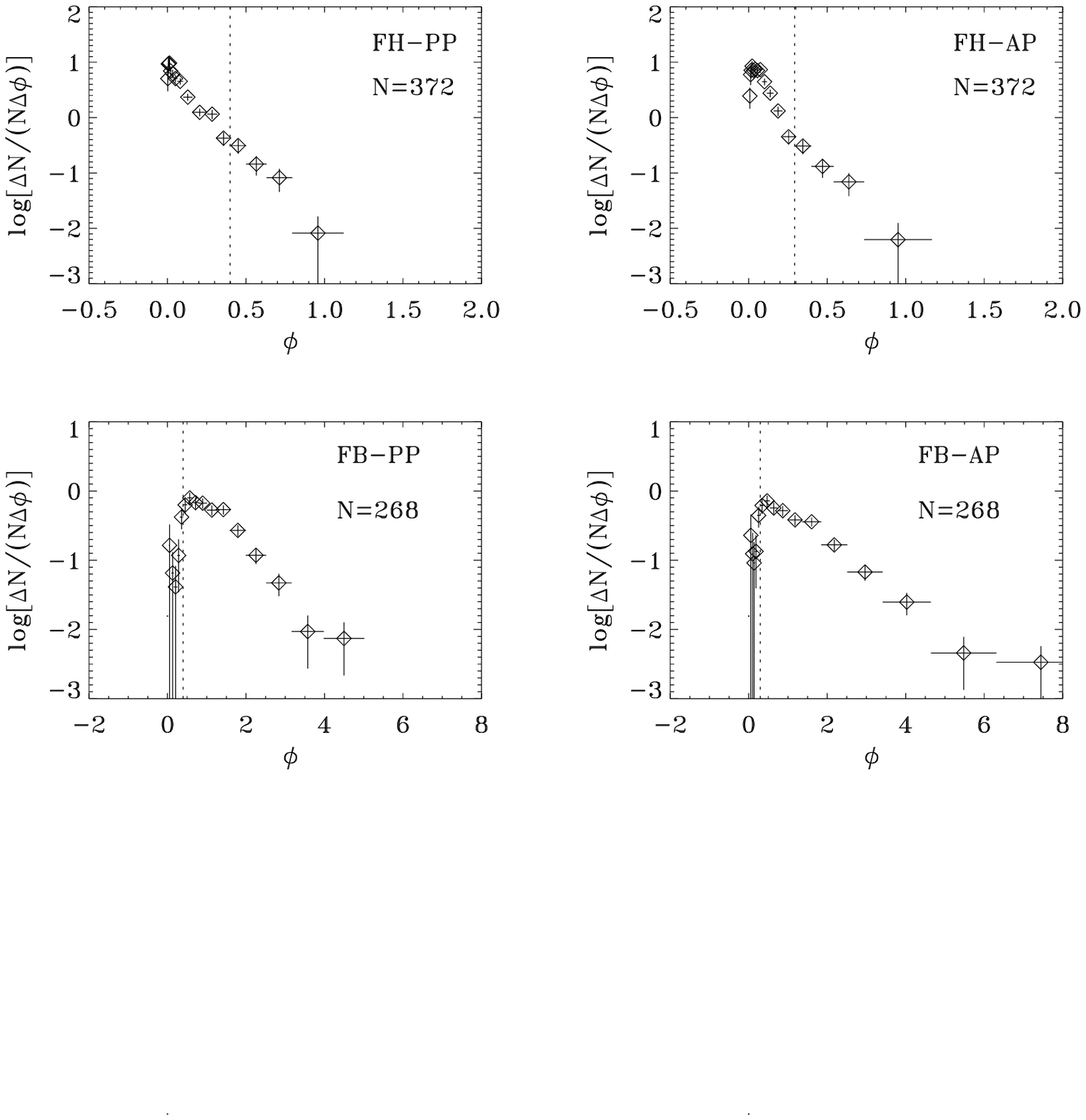,height=130mm,width=140mm}}
\caption[EGD]{The empirical, differential metallicity
distribution (EGD) in solar neighbourhood field halo
subdwarfs (top panels), plotted with regard to a sample
of 372 objects (Ryan \& Norris 1991), and in K-giant field
bulge stars in Baade's Window (bottom panels), plotted with
regard to a sample of 268 objects (Sadler et al. 1996),
both in presence (left panels) and in absence (right 
panels) of [O/Fe] plateau, respectively.
The dotted vertical line marks the transition from
halo (OH, YH) to bulge/disk (BD) morphological
type, [Fe/H]=$-$0.8.}
\label{f:EGD4}
\end{figure}

The related plots are shown in Fig.\,\ref{f:EGD4},
bottom left and bottom right, respectively.
Bottom panels represent the EGD deduced from a sample
of 372 halo subdwarfs in the solar neighbourhood,
studied by Ryan \& Norris (1991); for further details,
see C01.

\section{Inferred metallicity distribution in the
Galactic spheroid} \label{EGDs}

The Galactic spheroid may safely be conceived as
made of three main subsystems, namely: (i) globular
clusters, (ii) field halo stars, and (iii) field
bulge stars.   Let us assume a bulge mass, $M_B=
10^{10}{\rm m}_\odot$ (e.g., Wyse \& Gilmore 1992; Kent
1992; Prantzos \& Silk 1998), a (baryonic) halo
mass, $M_H=10^9{\rm m}_\odot$ (e.g., Carney, Latham \&
Laird 1990; Wyse \& Gilmore 1992), and a ratio
of globular clusters to halo mass, $M_{GC}/M_H=
1/100$ (e.g., Li \& Burstein 2003), where the
above values of bulge and halo mass may safely
be used including or not globular clusters.
Accordingly, the mass of the Galactic spheroid
and the fractional mass of related subsystems
read:
\begin{leftsubeqnarray}
\slabel{seq:Msa}
&& M=M_{\rm GC}+M_{\rm FH}+M_{\rm FB}=(0.001+0.1+1)~10^
{10}{\rm m}_\odot=1.101~{\rm m}_\odot~~; \\
\slabel{seq:Msb}
&& \frac{M_{\rm GC}}M=\frac1{1101}=0.0009~~; \\
\slabel{seq:Msc}
&& \frac{M_{\rm FH}}M=\frac{100}{1101}=0.0908~~; \\
\slabel{seq:Msd}
&& \frac{M_{\rm FB}}M=\frac{1000}{1101}=0.9083~~;
\label{seq:Ms}
\end{leftsubeqnarray}
where the indices, GC, FH, FB, mean globular
clusters, field halo stars, and field bulge
stars, respectively.

Let $N$ be the total number of long-lived
(i.e. life time longer than the age of the
Galaxy) stars in the Galactic spheroid,
and $\Delta N$ the number of long-lived
stars within a selected metallicity bin.
The relative frequency, $\Delta N/N$,
reads:
\begin{equation}
\label{eq:Ns}
\frac{\Delta N}N=\frac{N_{\rm GC}}N\frac
{\Delta N_{\rm GC}}{N_{\rm GC}}+\frac{N_
{\rm FH}}N\frac{\Delta N_{\rm FH}}{N_
{\rm FH}}+\frac{N_{\rm FB}}N\frac
{\Delta N_{\rm FB}}{N_{\rm FB}}~~;
\end{equation}
where $\Delta N_{\rm XY}/N_{\rm XY}$ is
the relative frequency belonging to XY
subsystem, XY = GC, FH, FB, with regard
to the selected metallicity bin.   Samples
related to different subsystems may be
(and in fact are) made of different objects,
provided each sample is statistically
significant.

With regard to field (halo and bulge) stars,
let us define a selected spectral class of
long-lived stars as belonging to a selected
mass range, $m_1\le m\le m_2$, and suppose
it does not significantly depend on the
evolution, as in disk stars (Rocha-Pinto
\& Maciel 1997).   Then the fractional number
of stars belonging to an assigned mass range,
equals the fractional number of stars
belonging to the whole mass domain, provided
the initial mass function (IMF) did not
significantly change within the Galactic
spheroid.   Accordingly, the following
relation holds:
\begin{equation}
\label{eq:NXY}
\left(\frac{N_{\rm XY}}N\right)_{m_1,m_2}=
\frac{N_{\rm XY}}N~~\qquad{\rm XY}={\rm GC},~
{\rm FH},~{\rm FB}~~;
\end{equation}
and, on the other hand:
\begin{equation}
\label{eq:MXY}
\frac{N_{\rm XY}}N=\frac{\overline{m}_{\rm XY}
N_{\rm XY}}{\overline{m}N}=\frac{M_{\rm XY}}M~~;
\end{equation}
where $\overline{m}_{\rm XY}$ and $\overline{m}$
are the mean mass of long-lived stars in XY
subsystem and in the Galactic spheroid, respectively,
and $\overline{m}_{\rm XY}=\overline{m}$
owing to the assumption of universal IMF.

The combination of Eqs.\,(\ref{eq:Ns}),
(\ref{eq:NXY}), and (\ref{eq:MXY}) yields: 
\begin{equation}
\label{eq:NMs}
\frac{\Delta N}N=\frac{M_{\rm GC}}M\frac
{\Delta N_{\rm GC}}{N_{\rm GC}}+\frac{M_
{\rm FH}}M\frac{\Delta N_{\rm FH}}{N_
{\rm FH}}+\frac{M_{\rm FB}}M\frac
{\Delta N_{\rm FB}}{N_{\rm FB}}~~;
\end{equation}
and the related uncertainty is obtained
using the standard formula of linear
propagation of errors together with
evaluation of Poisson errors (e.g.,
Ryan \& Norris 1991), $\Delta(\Delta N_
{\rm XY})=(\Delta N_{\rm XY})^{1/2}$.
The result is:
\begin{equation}
\label{eq:DNMs}
\Delta\frac{\Delta N}N=\frac{M_{\rm GC}}M\frac
{(\Delta N_{\rm GC})^{1/2}}{N_{\rm GC}}+\frac{M_
{\rm FH}}M\frac{(\Delta N_{\rm FH})^{1/2}}{N_
{\rm FH}}+\frac{M_{\rm FB}}M\frac
{(\Delta N_{\rm FB})^{1/2}}{N_{\rm FB}}~~;
\end{equation}
which may explicitly be calculated using
Eqs.\,(\ref{seq:Ms}) and the data listed
in Tab.\,\ref{t:E149}, Tabs.\,1-2 in C01,
and Tab.\,\ref{t:B262}, concerning globular
clusters, field halo stars, and field bulge
stars, respectively.

The EGD in stars of the Galactic spheroid
results from the combination of Eqs.\,(\ref
{eq:psi}), (\ref{seq:psier}), and (\ref{eq:NMs}),
as:
\begin{leftsubeqnarray}
\slabel{eq:psisa}
&& \psi=\log\left[\frac{M_{\rm GC}}M\frac
{\Delta N_{\rm GC}}{N_{\rm GC}\Delta\phi}+\frac{M_
{\rm FH}}M\frac{\Delta N_{\rm FH}}{N_
{\rm FH}\Delta\phi}+\frac{M_{\rm FB}}M\frac
{\Delta N_{\rm FB}}{N_{\rm FB}\Delta\phi}
\right]~~; \\
\slabel{eq:psisb}
&& \Delta^\mp\psi=\left\vert\log\left[1\mp
\frac{\Delta(\Delta N/N)}{\Delta N/N}\right]
\right\vert~~; \\
\slabel{eq:psisc}
&& \psi^\mp=\log\sum_{XY}\left[\frac{M_{\rm XY}}M\frac
{\Delta N_{\rm XY}\mp(\Delta N_{\rm XY})^{1/2}}{N_
{\rm XY}\Delta\phi}\right]~~;
\label{seq:psis}
\end{leftsubeqnarray}
where XY = GC, FH, FB.

\begin{figure}[t]
\centerline{\psfig{file=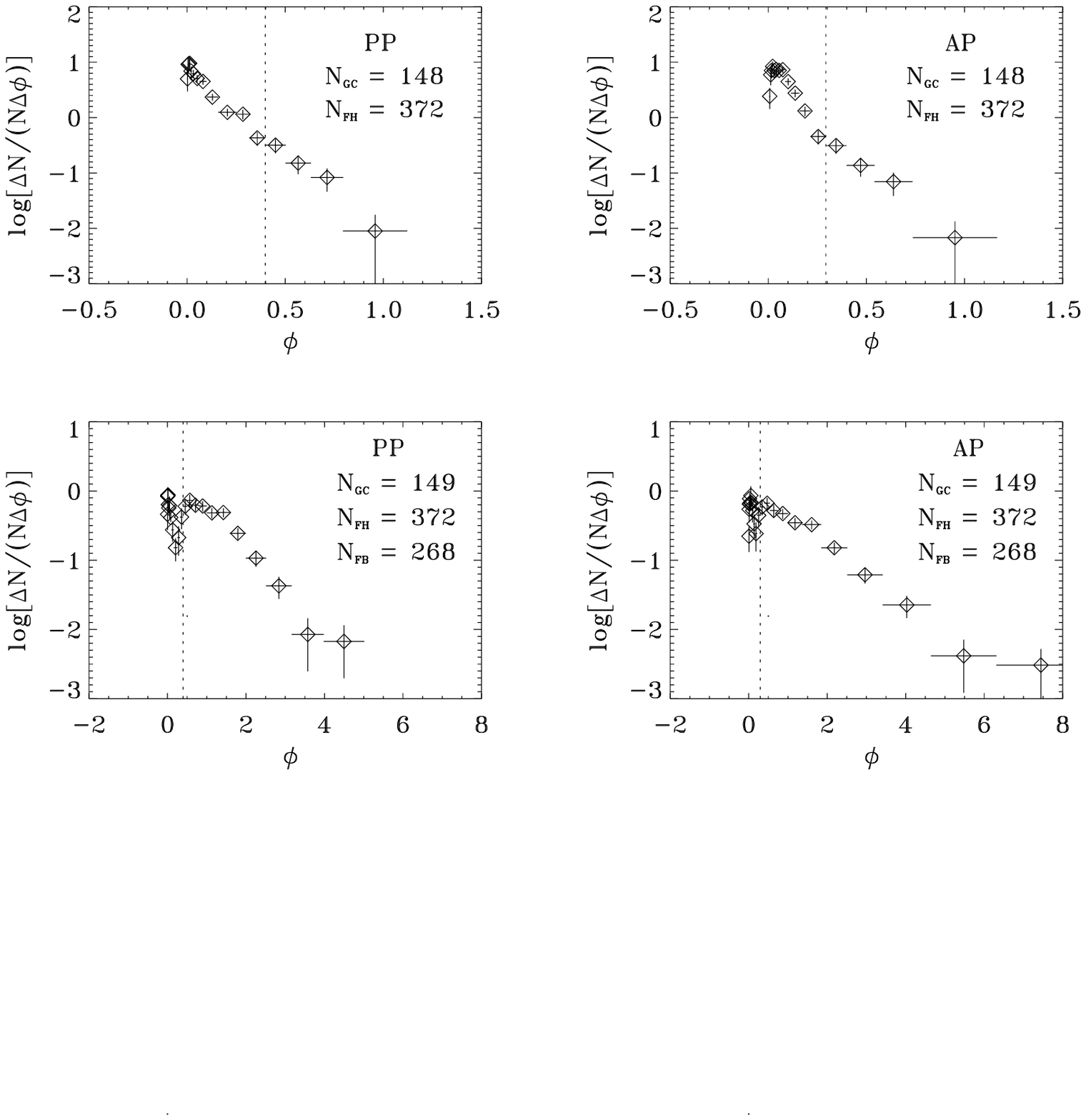,height=130mm,width=140mm}}
\caption[EGD]{The empirical, differential metallicity
distribution (EGD) in the light Galactic spheroid
(globular clusters and field halo stars; top panels),
and in the massive Galactic
spheroid (globular clusters, field halo stars, and field
bulge stars; bottom panels),
both in presence (left panels) and in absence (right 
panels) of [O/Fe] plateau, respectively.
The distribution has been deduced using Eqs.\,(\ref
{eq:NMs}), (\ref{eq:DNMs}), and (\ref{seq:psis}),
from a sample of 148 (top panels) or 149 (bottom panels)
globular clusters (Mackey \& van den Bergh 2005),
a sample of 372 field halo subdwarfs (Ryan \& Norris
1991), and a sample of 268 K-giant field bulge stars
(Sadler et al. 1996).
The dotted vertical line marks the transition from
halo (OH, YH) to bulge/disk (BD) morphological
type, [Fe/H]=$-$0.8.}
\label{f:EGD3}
\end{figure}

As a first application of  Eqs.\,(\ref
{eq:NMs}), (\ref{eq:DNMs}), and (\ref{seq:psis}),
let us take into consideration the light Galactic
spheroid, made of globular clusters and field
halo stars.   To this aim, the globular
cluster sample (Mackey \& van den Bergh 2005)
has been reduced to $N=148$ with the exclusion
of Liller I owing to its high metal content,
[Fe/H]=0.22, which exceeds the maximum
metallicity in solar neighbourhood halo
subdwarf sample (Ryan \& Norris 1991)
used here.   The resulting EGD is listed
in Tab.\,\ref{t:EH} both in presence and
in absence of [O/Fe] plateau.
\begin{table}
\caption[par]{The empirical, differential
metallicity distribution (EGD) in the light
Glalctic spheroid (globular clusters and field halo 
stars), deduced from a sample of 148 globular
clusters (Mackey \& van den Bergh 2005) and
a sample of 372 field halo subdwarfs (Ryan
\& Norris 1991), both in presence (PP)
and in absence (AP) of [O/Fe] 
plateau.}
\label{t:EH}
\begin{center}
\begin{tabular}{rrrrrrrr}
\multicolumn{2}{c|}{PP}
&\multicolumn{2}{c|}{AP} \\
\hline\noalign{\smallskip}
\multicolumn{1}{c}{$\phi$} & \multicolumn{1}{c}{$\phantom{0}\psi$} &
\multicolumn{1}{c}{$\phi$} & \multicolumn{1}{c}{$\phantom{0}\psi$} &
\multicolumn{1}{c}{$\Delta^-\psi$} & \multicolumn{1}{c}{$\Delta^+\psi$} &
\multicolumn{1}{c}{$\frac{\Delta N}N$} & \multicolumn{1}{c}{$\Delta\frac
{\Delta N}N$}  \\
\noalign{\smallskip}
\hline\noalign{\smallskip}
0.958 & $-$2.048 & 0.951 & $-$2.167 & 1.336 & 0.291 &
0.003 & 0.003 \\
0.713 & $-$1.080 & 0.638 & $-$1.156 & 0.258 & 0.161 &
0.014 & 0.006 \\
0.566 & $-$0.821 & 0.470 & $-$0.864 & 0.202 & 0.137 &
0.020 & 0.007 \\
0.450 & $-$0.496 & 0.345 & $-$0.505 & 0.148 & 0.110 &
0.033 & 0.009 \\
0.357 & $-$0.367 & 0.254 & $-$0.342 & 0.142 & 0.107 &
0.035 & 0.010 \\
0.284 & 0.062 & 0.187 & 0.120 & 0.092 & 0.076 &
0.075 & 0.014 \\
0.205 & 0.096 & 0.137 & 0.441 & 0.072 & 0.062 &
0.116 & 0.018 \\
0.129 & 0.371 & 0.101 & 0.650 & 0.066 & 0.057 &
0.137 & 0.019 \\
0.081 & 0.653 & 0.074 & 0.865 & 0.059 & 0.052 &
0.166 & 0.021 \\
0.051 & 0.707 & 0.055 & 0.852 & 0.071 & 0.061 &
0.118 & 0.018 \\
0.032 & 0.792 & 0.040 & 0.870 & 0.082 & 0.069 &
0.091 & 0.016 \\
0.020 & 0.840 & 0.030 & 0.852 & 0.100 & 0.081 &
0.064 & 0.013 \\
0.013 & 0.980 & 0.022 & 0.925 & 0.107 & 0.086 &
0.056 & 0.012 \\
0.008 & 0.972 & 0.016 & 0.850 & 0.141 & 0.106 &
0.035 & 0.010 \\
0.005 & 0.961 & 0.012 & 0.773 & 0.189 & 0.131 &
0.021 & 0.008 \\
0.002 & 0.700 & 0.007 & 0.385 & 0.228 & 0.149 &
0.016 & 0.007 \\
\noalign{\smallskip}
\hline
\end{tabular}
\end{center}
\end{table}
The related plots are shown in Fig.\,\ref{f:EGD3}
(top panels), left and right, respectively.

A second application concerns the massive
Galactic spheroid, including globular clusters,
field halo stars, and field bulge stars.   The
resulting EGD is listed 
in Tab.\,\ref{t:EHB} both in presence and
in absence of [O/Fe] plateau.
\begin{table}
\caption[par]{The empirical, differential
metallicity distribution (EGD) in the massive
Galactic spheroid (globular clusters, field halo 
stars, and field bulge stars), deduced from a 
sample of 149 globular
clusters (Mackey \& van den Bergh 2005),
a sample of 372 field halo subdwarfs (Ryan
\& Norris 1991), and a sample of 268 field
K-giant bulge stars (Sadler et al. 1996), 
both in presence (PP) and in absence (AP) 
of [O/Fe] plateau.}
\label{t:EHB}
\begin{center}
\begin{tabular}{rrrrrrrr}
\multicolumn{2}{c|}{PP}
&\multicolumn{2}{c|}{AP} \\
\hline\noalign{\smallskip}
\multicolumn{1}{c}{$\phi$} & \multicolumn{1}{c}{$\phantom{0}\psi$} &
\multicolumn{1}{c}{$\phi$} & \multicolumn{1}{c}{$\phantom{0}\psi$} &
\multicolumn{1}{c}{$\Delta^-\psi$} & \multicolumn{1}{c}{$\Delta^+\psi$} &
\multicolumn{1}{c}{$\frac{\Delta N}N$} & \multicolumn{1}{c}{$\Delta\frac
{\Delta N}N$}  \\
\noalign{\smallskip}
\hline\noalign{\smallskip}
4.496 & $-$2.172 & 7.443 & $-$2.515 & 0.533 & 0.232 & 0.007 & 0.005 \\
3.572 & $-$2.072 & 5.476 & $-$2.381 & 0.533 & 0.232 & 0.007 & 0.005 \\
2.837 & $-$1.370 & 4.028 & $-$1.646 & 0.189 & 0.131 & 0.028 & 0.010 \\
2.254 & $-$0.969 & 2.963 & $-$1.211 & 0.125 & 0.097 & 0.055 & 0.014 \\
1.790 & $-$0.611 & 2.180 & $-$0.820 & 0.089 & 0.074 & 0.100 & 0.019 \\
1.422 & $-$0.310 & 1.604 & $-$0.486 & 0.069 & 0.060 & 0.159 & 0.023 \\
1.129 & $-$0.316 & 1.180 & $-$0.458 & 0.080 & 0.068 & 0.125 & 0.021 \\
0.897 & $-$0.217 & 0.868 & $-$0.326 & 0.079 & 0.067 & 0.125 & 0.021 \\
0.713 & $-$0.205 & 0.638 & $-$0.281 & 0.091 & 0.075 & 0.102 & 0.019 \\
0.566 & $-$0.134 & 0.470 & $-$0.176 & 0.095 & 0.078 & 0.095 & 0.019 \\
0.450 & $-$0.221 & 0.341 & $-$0.230 & 0.122 & 0.095 & 0.062 & 0.015 \\
0.357 & $-$0.376 & 0.254 & $-$0.352 & 0.173 & 0.127 & 0.034 & 0.011 \\
0.284 & $-$0.673 & 0.187 & $-$0.615 & 0.259 & 0.161 & 0.014 & 0.006 \\
0.205 & $-$0.819 & 0.137 & $-$0.474 & 0.195 & 0.134 & 0.014 & 0.005 \\
0.129 & $-$0.561 & 0.101 & $-$0.282 & 0.171 & 0.122 & 0.016 & 0.005 \\
0.081 & $-$0.384 & 0.074 & $-$0.172 & 0.059 & 0.052 & 0.015 & 0.002 \\
0.051 & $-$0.211 & 0.055 & $-$0.066 & 0.192 & 0.132 & 0.014 & 0.005 \\
0.032 & $-$0.245 & 0.040 & $-$0.167 & 0.082 & 0.069 & 0.008 & 0.001 \\
0.020 & $-$0.197 & 0.030 & $-$0.185 & 0.100 & 0.081 & 0.006 & 0.001 \\
0.013 & $-$0.057 & 0.022 & $-$0.112 & 0.107 & 0.086 & 0.005 & 0.001 \\
0.008 & $-$0.065 & 0.016 & $-$0.187 & 0.141 & 0.106 & 0.003 & 0.001 \\
0.005 & $-$0.076 & 0.012 & $-$0.265 & 0.189 & 0.131 & 0.002 & 0.001 \\
0.002 & $-$0.337 & 0.007 & $-$0.653 & 0.228 & 0.149 & 0.001 & 0.001 \\
\noalign{\smallskip}
\hline
\end{tabular}
\end{center}
\end{table}
The related plots are shown in Fig.\,\ref{f:EGD3}
(bottom panels), left and right, respectively.

The EGD in the Galactic spheroid is similar to
its counterpart related to globular cluster
subsystem, Fig.\,\ref{f:EGD2}, which is bimodal
with the occurrence of two distinct maxima,
and a minimum soon before the transition from
halo to bulge/disk morphological type,
represented by a dotted vertical line.
It is apparent that (closed or open) simple
models of chemical evolution cannot provide
a satisfactory explanation to a bimodal EGD,
and a different model is needed.

Accordingly, let us suppose the halo and the
bulge underwent distinct chemical evolutions,
calculate the related theoretical differential
metallicity distribution (hereafter referred 
to as TGD) in a selected spectral class of
long-lived stars, the resulting TGD, and
compare with its empirical counterpart.   
To this aim, simple models implying both
homogeneous and inhomogeneous mixing shall
be used, as in C01.

\section{Homogeneous simple models}
\label{osmo}

With regard to homogeneous simple models
with star formation inhibiting gas (e.g.,
Hartwick 1976; C00; C01), the TGD is
represented as a straight line (e.g., 
Pagel 1989; C00; C01):
\begin{equation}
\label{eq:psil}
\psi(\phi)=\log\frac{\diff N}{N\diff\phi}=
a\phi+b~~;
\end{equation}
and the explicit expression of the coefficients,
$a$ and $b$, reads:
\begin{lefteqnarray}
\label{eq:a}
&& a=-\frac1{\ln10}\frac{{\rm O}_\odot}{\hat{p}^{\prime\prime}}~; \\
\label{eq:b}
&& b=\log\left(\frac{\mu_o}{\mu_o-\mu_f}\frac{{\rm O}_\odot}
{\hat{p}^{\prime\prime}}\right)-a\phi_o~;
\end{lefteqnarray}
where ${\rm O}_\odot$ is the solar oxygen abundance,
$\hat{p}^{\prime\prime}$ the effective (oxygen)
yield, $\mu$ the (allowing star formation) gas mass
fraction, and the indices, $o$ and $f$, denote
the beginning and the end of evolution, respectively.

The oxygen abundance, O, may be related, to a
good extent, to the gas mass fraction, $\mu$, as
(e.g., C00; C01):
\begin{lefteqnarray}
\label{eq:Oyz}
&& {\rm O}-{\rm O}_0=-\hat{p}^{\prime\prime}\ln
\frac{\mu}{\mu_0}~~; \\
\label{eq:ypy}
&& \hat{p}^{\prime\prime}=\frac{\hat{p}}{1+\kappa}~~;
\end{lefteqnarray}
where $\hat{p}$, is the real (oxygen) yield and
$\kappa$ is the ratio of (inhibiting star formation)
gas mass fraction to long-lived star and
remnant mass fraction.   In addition, the following
assumptions have been made: (i) instantaneous
recycling; (ii) universal power-law IMF; and
(iii) gas inhibition from forming stars at a
rate proportional to the star formation rate.
For further details see e.g., Hartwick (1976);
Pagel (1989); C00; C01.

A deeper analysis shows that, given a simple model
with star formation inhibiting gas, long-lived
lower stellar mass limit, $m_{mf}$, and inhibition
parameter, $\kappa$, it is equivalent to a simple
model with same values of independent parameters
except the above mentioned two, $(m_{mf})_1\le m_
{mf}$ and $\kappa_1=0\le\kappa$; and vice versa.
Accordingly, the related yield, $\hat{p}_1$, has
the same value as the effective yield, $\hat{p}^
{\prime\prime}$, expressed by Eq.\,(\ref{eq:ypy}).
For an exhaustive discussion, see C01.

Simple models with star formation inhibiting
gas (Hartwick 1976)
can be generalized to negative values of the
inhibition parameter, $\kappa<0$, where star
formation is enhanced by additional gas with
same composition as the pre-existing one, at
a rate:
\begin{equation}
\label{eq:desf}
\frac{\diff D}{\diff t}=\alpha\kappa\frac{\diff
S}{\diff t}~~;\qquad\kappa<0~~;
\end{equation}
which corresponds to a total amount:
\begin{equation}
\label{eq:iesf}
D(t)-D_o=\alpha\kappa[S(t)-S_o]=\kappa[s(t)-s_o]~~;
\qquad\kappa<0~~;
\end{equation}
where $D$, $S$, and $s$ represent the mass fraction 
in additional gas, stars formed, and long-lived stars
and remnants, respectively, and $\alpha$ is the mass
fraction of a star generation which remains locked
up in long-lived stars and remnants.

The additional gas may be conceived as inflowing
from outside, with same composition as the 
pre-existing gas, and entirely turned into stars.
Accordingly, the mass fraction with respect to
the original system, $\mu+s$, reads:
\begin{equation}
\label{eq:musD}
\mu(t)+s(t)=1-D(t)~~;\qquad\kappa<0~~;
\end{equation}
and the combination of Eqs.\,(\ref{eq:iesf})
and (\ref{eq:musD}) yields:
\begin{lefteqnarray}
\label{eq:smu}
&& s-s_o=\frac{\mu_0-\mu}{1+\kappa}~~; \\
\label{eq:Dmu}
&& D-D_o=\frac{\kappa(\mu_o-\mu)}{1+\kappa};~~
\end{lefteqnarray}
finally, the related distribution of metal
abundance in long-lived stars can be determined
following a standard procedure (e.g., Pagel \&
Patchett 1975; Caimmi 1981), as done in C01.
In fact, all the considerations made in C01
(Appendix C) are independent of the sign of
the inhibition parameter, $\kappa$, and Theorem
1 therein may be generalized in the following
way.
\begin{trivlist}
\item[\hspace\labelsep{\bf Theorem}] \sl
Given a simple model of chemical evolution with
star formation inhibiting gas,
long-lived (i.e. life time longer than the
age of the system) lower stellar mass limit,
$m_{mf}$, and inhibition parameter, $\kappa$
(where positive and negative values correspond
to star formation inhibiting and enhancing gas,
respectively),
it is equivalent to any model of the same kind
and with same values of parameters, except the
above mentioned two, $(m_{mf})_n$ and $\kappa_n$,
which are defined by the relations:
\begin{leftsubeqnarray}
\slabel{eq:teo1a}
&& (m_{mf})_1\le(m_{mf})_n\le m_{mf}~~;\qquad
\kappa\ge0~~; \\
\slabel{eq:teo1b}
&& (m_{mf})_n\le(m_{mf})_1\le m_{mf}~~;\qquad
-1<\kappa\le0~~;
\label{seq:teo1}
\end{leftsubeqnarray}
\begin{lefteqnarray}
\label{eq:teo2}
&& \frac{1+\kappa_n}{1+\kappa}=\frac{\hat{p}_n}
{\hat{p}}~~;
\end{lefteqnarray}
where $(m_{mf})_1$ is related to $\kappa_n=0$,
i.e. star formation neither inhibiting nor
enhancing gas.
\end{trivlist}
Aiming to an application of the above results
to the chemical evolution of the Galactic spheroid,
the value of the real normalized yield, $\hat{p}/
{\rm O}_\odot$, the long-lived star and remnant mass
fraction in a star generation, $\alpha$, and the
lower stellar mass limit, $m_{mf}$, shall be
taken from a fit to the EGD in the disk solar
neighbourhood (C00) with regard to a solar
oxygen abundance, ${\rm O}_\odot=0.0056$ (C01).

Two extreme values of the power-law IMF exponent,
$p$, shall be considered, namely: (i) $p=2.9$,
which is a fit to the IMF determined by
Scalo (1986) or Miller \& Scalo (1979),
for $m\appgeq {\rm m}_\odot$, and provides a good
approximation also in terms of oxygen production
(Wang \& Silk 1993); in fact, a different
model (with respect to the current one) for
the chemical evolution of the Galactic halo
also requires a Miller-Scalo IMF to fit the
data (Lu et al. 2001), and (ii) $p=2.35$, which
coincides with the IMF determined by Salpeter
(1955). Different fits to the EGD in the Galactic
spheroid, with respect to the solar neighbourhood,
would imply star formation inhibiting or enhancing
gas.
For deeper insight into the model, see C00, C01.

Input parameters which remain fixed are listed in
Tab.\,\ref{t:infi}, where the indices, 2.9 and 2.35,
denote the value of the power-law IMF exponent
used in computing the corresponding quantities.
\begin{table}
\caption[pahd]{Values of input parameters
of simple homogeneus models with star
formation inhibiting or enhancing gas.    The
indices, 2.9 and 2.35,
denote the value of the power-law IMF exponent
used in computing the corresponding quantities.
The indices, H and B, denote halo and bulge 
field star subsystem, respectively.}
\label{t:infi}
\begin{center}
\begin{tabular}{ll}
\hline\noalign{\smallskip}
$\hat{p}/{\rm O}_\odot\cdot10$ & 7.3722 \\
$(\widetilde{m}_{mf})_{2.9}\cdot10$ & 3.4235 \\
$(\widetilde{m}_{mf})_{2.35}\cdot10^3$ & 6.9136 \\
$\alpha_{2.9}\cdot10$ & 7.3666 \\
$\alpha_{2.35}\cdot10$ & 8.9104 \\
$\mu_o$ & 1 \\
$s_o$ & 0 \\
$D_o$ & 0 \\
$\phi_{oH}\cdot10^3$ & 1 \\
$\phi_{oB}$ & 0.20 \\
$\phi_{fH}$ & 1 \\
$\phi_{fB}$ & 5.5 \\
${\rm O}_\odot\cdot10^3$ & 5.6 \\
\noalign{\smallskip}
\hline
\end{tabular}
\end{center}
\end{table}
The initial oxygen abundance assumed for field
halo stars, $\phi_{oH}=0.001$, is consistent
with a lower limit deduced from theoretical
considerations, to allow pop.\,II star formation
(Bromm \& Loeb 2003).

Different models can be obtained by changing a
single remaining input parameter, which can be
chosen as the normalized effective yield, $\hat
{p}^{\prime\prime}/{\rm O}_\odot$, or the slope of
the TGD, $a$, conform to Eqs.\,(\ref{eq:psil})
and (\ref{eq:a}).

A few cases are listed in Tab.\,\ref{t:HB},
related to halo (H) and bulge (B) field
stars, respectively, where $\overline{\phi}$
represents the mean oxygen abundance
(normalized to the solar value) of stars
at the end of evolution.
\begin{table}
\caption[pahd]{Values of parameters related
to homogeneous simple models with star formation
inhibiting or enhancing gas, corresponding
to linear fits to the empirical differential
metallicity distribution (EGD) in halo (H) and
bulge (B) field stars, plotted in Fig.\,\ref
{f:EGD4}.   The mean oxygen abundance
(normalized to the solar value) of stars
at the end of evolution is denoted as $\overline
{\phi}$.   Positive
and negative $\kappa$ values mean star formation
inhibiting and enhancing gas, respectively.
Positive and negative $D$ values mean star formation
inhibiting gas and stars formed from additional
gas with same composition as the pre-existing gas,
respectively.  Mass fractions are normalized to
the initial mass.}
\label{t:HB}
\begin{center}
\begin{tabular}{crrrr}
\multicolumn{1}{c}{} & \multicolumn{1}{c}{H1} &
\multicolumn{1}{c}{H2} & \multicolumn{1}{c}{B1} &
\multicolumn{1}{c}{B2} \\
\hline\noalign{\smallskip}
$\hat{p}^{\prime\prime}/{\rm O}_\odot$ & 1.3363~E$-$1 & 9.2288~E$-$2 &
1.0857~E$-$0 & 8.1430~E$-$1 \\
$a$ & $-$3.2500~E$-$0 & $-$4.7059~E$-$0 & $-$4.0000~E$-$1 & $-$5.3333~E$-$1 \\
$b$ & 8.7771~E$-$1 & 1.0396~E$-$0 & 1.1048~E$-$1 & 2.3465~E$-$1 \\
$\kappa$ & 4.5169~E$-$0 & 6.9883~E$-$0 & $-$3.2099~E$-$1 & $-$9.4658~E$-$2 \\
$\mu_f$ & 5.6657~E$-$4 & 1.9899~E$-$5 & 7.5858~E$-$3 & 1.4905~E$-$3 \\
$s_f$ & 1.8116~E$-$1 & 1.2518~E$-$1 & 1.4616~E$-$0 & 1.1029~E$-0$ \\
$D_f$ & 8.1828~E$-$1 & 8.7480~E$-$1 & $-$4.6915~E$-$1 & $-$1.0440~E$-$1 \\
$\overline{\phi}$ & 1.3406~E$-$1 & 9.3268~E$-$2 & 1.2452~E$-$0 &
1.0064~E$-$0 \\
\noalign{\smallskip}
\hline
\end{tabular}
\end{center}
\end{table}
Positive
and negative $\kappa$ values mean star formation
inhibiting and enhancing gas, respectively.
Positive and negative $D$ values mean star formation
inhibiting gas and stars formed from additional
gas with same composition as the pre-existing gas,
respectively.

It is apparent that halo and bulge
models demand star formation inhibiting and
enhancing gas, respectively, to maintain (i)
universal power-law IMF, and (ii) real normalized
yield, $\hat{p}/{\rm O}_\odot$, unchanged with respect
to a value deduced from an acceptable fit to the
EGD in the disk solar neighbourhood (C00; C01).
More precisely, about 80-90\% of the initial
halo gas has to be inhibited from forming stars
and, on the other hand, about 10-50\% of the
initial bulge mass has to be added as gas
enhanced in forming stars, for providing an
acceptable fit to the related EGD.  It
results in a normalized effective yield
lower by about one order of magnitude in
the halo, with respect to the bulge.

\begin{figure}[t]
\centerline{\psfig{file=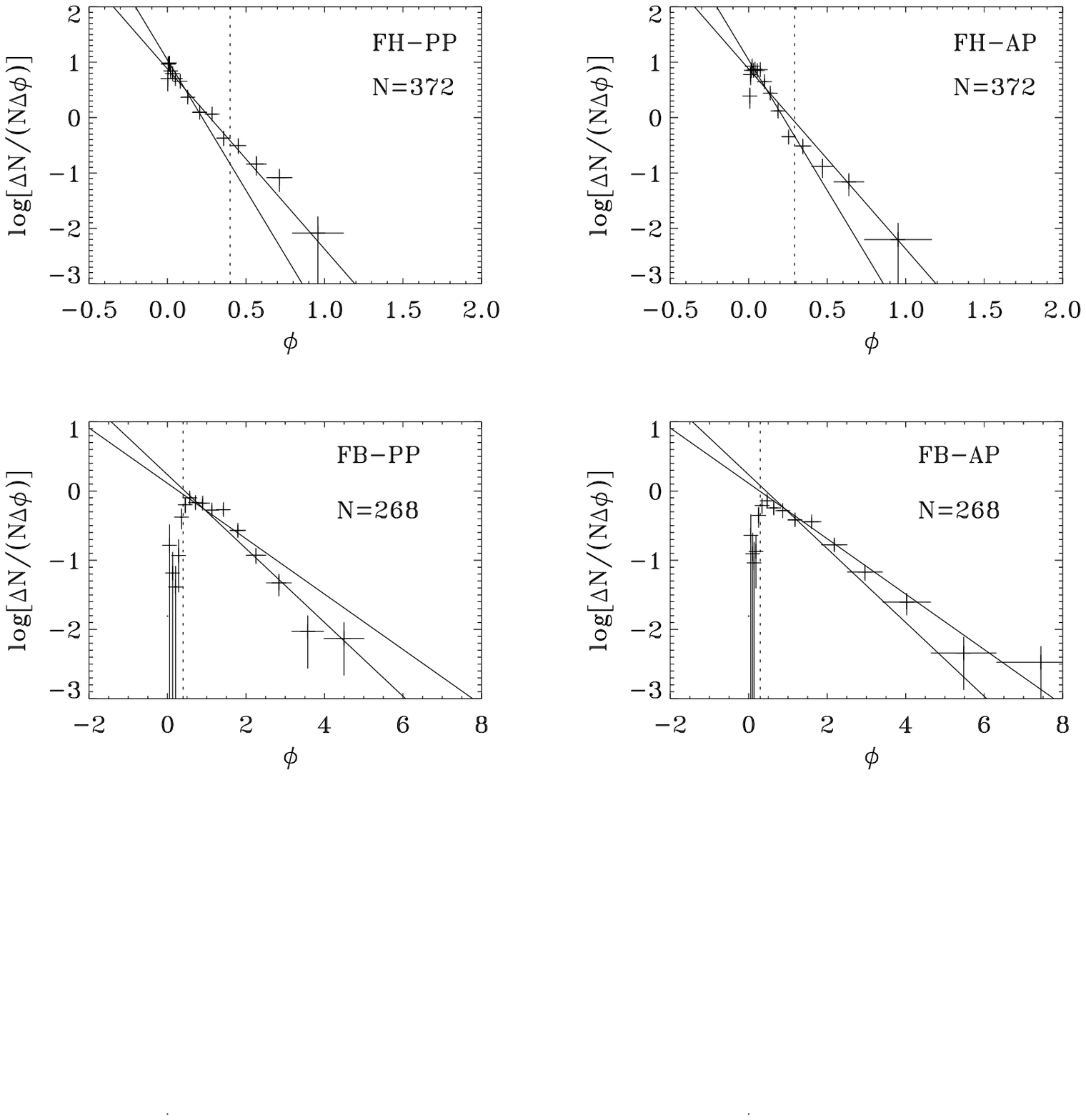,height=130mm,width=140mm}}
\caption[EGD]{Comparison between theoretical (TGD)
and empirical differential oxygen abundance
distribution (EGD) in halo (top panels) and bulge
(bottom panels) field stars, 
both in presence (left panels) and in absence (right 
panels) of [O/Fe] plateau, respectively.
The straight lines correspond to homogeneous
models, H1-H2 (top panels) and B1-B2 (bottom panels)
defined in Tabs.\,\ref{t:infi} and \ref{t:HB}.
Crosses represent the data and related
uncertainties, as in Fig.\,\ref{f:EGD4}.
The dotted vertical line marks the transition from
halo (OH, YH) to bulge/disk (BD) morphological
type in globular clusters, [Fe/H]=$-$0.8.}
\label{f:TGD4}
\end{figure}

The TGD deduced from models H1-H2 and B1-B2
is represented in Fig.\,\ref{f:TGD4} and
compared to the corresponding EGD in
connection with halo (top panels) and bulge
(bottom panels) field stars, both in
presence (left panels) and in absence
(right panels) of [O/Fe] plateau.

With regard to the halo, homogeneous simple
models in presence of star formation inhibiting
gas, provide an acceptable fit with values of
input parameters as listed in Tab.\,\ref
{t:infi} and $0.092<\hat{p}^{\prime\prime}/
{\rm O}_\odot<0.134$, both in presence and in
absence of [O/Fe] plateau.

With regard to the bulge, homogeneous simple
models in presence of star formation enhancing
gas, provide an acceptable fit with values of
input parameters as listed in Tab.\,\ref
{t:infi} and $0.81<\hat{p}^{\prime\prime}/
{\rm O}_\odot<1.09$, both in presence and in
absence of [O/Fe] plateau.

Further inspection of Fig.\,\ref{f:TGD4}
shows that, in both cases, there is a
slight deficiency (strenghtened in absence
of [O/Fe] plateau), in the number of stars
observed below a treshold, $\phi\approx
0.004$ or [Fe/H]$\approx-3.0$ with regard
to the halo, and $\phi\approx0.5$ or
[Fe/H]$\approx-0.6$ with regard to the
bulge.   In other words, a G-dwarf
problem seems to exist for both the
halo (Hartwick 1976; Prantzos 2003)
and the bulge (Ferreras et al. 2003).

\begin{figure}[t]
\centerline{\psfig{file=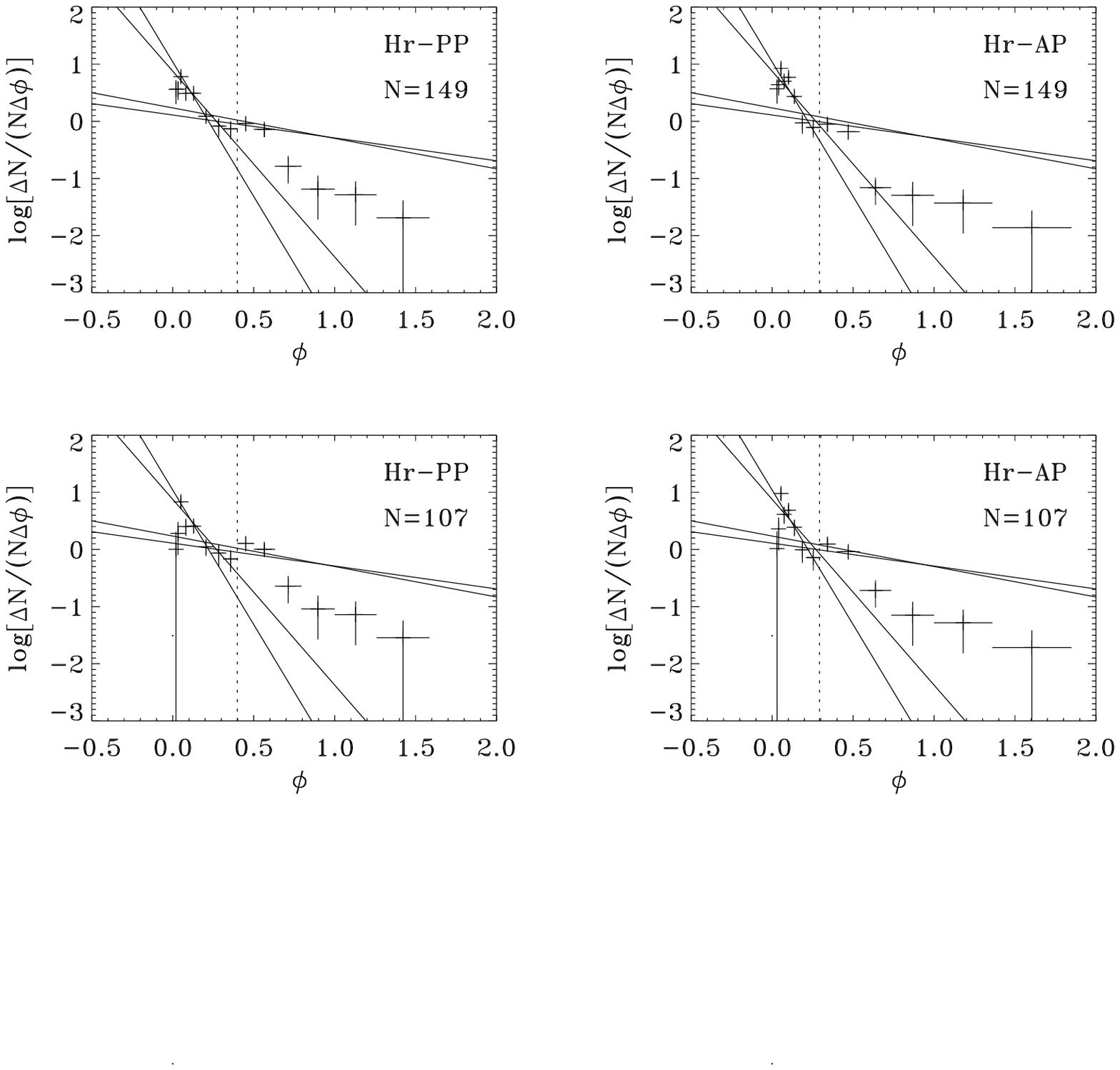,height=130mm,width=140mm}}
\caption[EGD]{Comparison between theoretical (TGD)
and empirical differential oxygen abundance
distribution (EGD) in globular clusters of
all morphological types (top panels) and with
only old halo and bulge/disk objects retained
(bottom panels), 
both in presence (left panels) and in absence
(right panels) of [O/Fe] plateau, respectively.
The straight lines correspond to homogeneous
models, H1-H2 (more inclined) and B1-B2 (less
inclined), respectively,
defined in Tabs.\,\ref{t:infi} and \ref{t:HB}.
Crosses represent the data and related
uncertainties, as in Fig.\,\ref{f:EGD2}.
The dotted vertical line marks the transition from
halo (OH, YH) to bulge/disk (BD) morphological
type in globular clusters, [Fe/H]=$-$0.8.}
\label{f:TGD3}
\end{figure}

The TGD deduced from models H1-H2 and B1-B2
is represented in Fig\,\ref{f:TGD3} and
compared to the corresponding EGD in
connection with the complete sample $(N=149)$
of globular clusters considered above (top
panels), and a reduced sample $(N=107)$
with only old halo and bulge/disk objects
retained (bottom panels), both in
presence (left panels) and in absence
(right panels) of [O/Fe] plateau.
With regard to halo clusters, homogeneous simple
models with star formation inhibiting gas,
provide an acceptable fit leaving aside the
occurrence of a G-dwarf problem.  Accordingly,
field halo stars and halo globular clusters
underwent similar chemical evolution, which
is consistent with recent results on the
comparison of elemental abundance ratios
(Pritzl, Venn \& Irwin 2005).

With regard to bulge/disk clusters, homogeneous
simple models with star formation enhancing gas,
provide an acceptable fit below a treshold in
metal abundance, $\phi\approx0.6$ or [Fe/H]$
\approx0.5$, while a G-dwarf problem appears
when the treshold is exceeded.   On the other
hand, it is not the case for bulge field stars,
which implies that the formation of bulge/disk
metal-rich globular clusters was inhibited (in
the sense that proto-cluster stars turned into
field stars) and/or tidal disruption took place.
Accordingly, bulge (and possibly thick disk)
field stars and bulge/disk globular clusters
underwent similar chemical evolution.

Clusters within a restricted metallicity range,
$-1.2\appleq$[Fe/H]$\appleq-0.8$, are fitted by
both halo and bulge models, which allows to
shift on the left the transition from halo to
bulge/disk, from [Fe/H]$\approx-0.8$ to [Fe/H]
$\approx-1.2$, consistent with a restricted
range in age shown by metal-rich clusters,
[Fe/H]$\appgeq-1.0$, see Fig.\,\ref{f:MAMR}.

With regard to Galactic spheroid field stars, the
differential version of Eq.\,(\ref{eq:psisa})
reads:
\begin{leftsubeqnarray}
\slabel{eq:dpsia}
&& \psi=\log\frac{\diff N}{N\diff\phi}=\log\frac
{\diff N_H+\diff N_B}{N\diff\phi}\nonumber \\
&& \phantom{\psi}=\log\left[\frac{N_H}N\frac
{\diff N_H}{N_H\diff\phi}+\frac{N_B}N\frac
{\diff N_B}{N_B\diff\phi}\right]~~; \\
\slabel{eq:dpsib}
&& N_H=N_{FH}+(N_{GC})_H~~;\qquad
N_B=N_{FB}+(N_{GC})_B~~;
\label{seq:dpsi}
\end{leftsubeqnarray}
which includes the contribution from globular
clusters.   The combination of Eqs.\,(\ref
{eq:MXY}), (\ref{eq:psil}), and (\ref{seq:dpsi})
yields:
\begin{equation}
\label{eq:psti}
\psi=\log\left[\frac{M_H}M\exp_{10}(a_H\phi+b_H)+
\frac{M_B}M\exp_{10}(a_B\phi+b_B)\right]~~;
\end{equation}
where, in general, $\exp_ux=x^u$ and $\exp_{\rm e}
x={\rm e}^x$ according to the usual notation.
Then the TGD related to the Galactic spheroid
is expressed by Eq.\,(\ref{eq:psti}).

\begin{figure}[t]
\centerline{\psfig{file=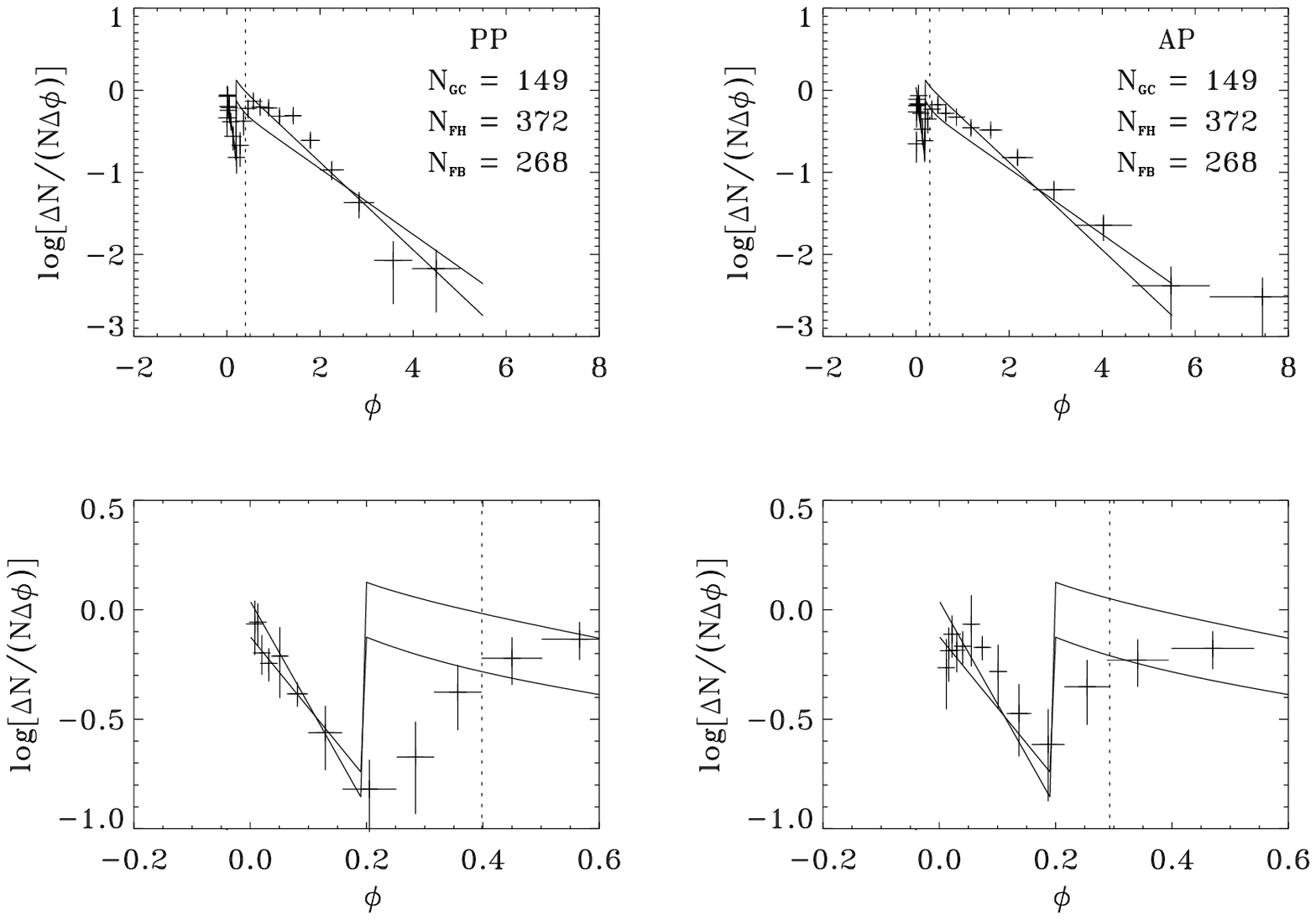,height=130mm,width=140mm}}
\caption[EGD]{Comparison between theoretical (TGD)
and empirical differential oxygen abundance
distribution (EGD) in the Galactic spheroid
(top panels) and zoomed for low oxygen abundance
(bottom panels), 
both in presence (left panels) and in absence (right 
panels) of [O/Fe] plateau, respectively.
The curves correspond to
models, H1, B1; H2, B2;
defined in Tabs.\,\ref{t:infi} and \ref{t:HB}
and combined via Eq.\,(\ref{eq:psti}).
Crosses represent the combination of the data and
related uncertainties via Eqs.\,(\ref{seq:psis}),
as in Fig.\,\ref{f:EGD3} (bottom panels).
The dotted vertical line marks the transition from
halo (OH, YH) to bulge/disk (BD) morphological
type in globular clusters, [Fe/H]=$-$0.8.}
\label{f:TGD5}
\end{figure}

Two alternatives shall be considered, $S_1$ and
$S_2$, according if the coefficients, $a$ and $b$,
are taken from cases H1, B1, and H2, B2,
of Tab.\,\ref{t:HB}, respectively, and $M_H/M=0.1$,
$M_B/M=0.9$, are assumed to a good extent.
The resulting TGD is plotted in Fig.\,\ref
{f:TGD5} and compared to its empirical counterpart,
represented in Fig.\,\ref{f:EGD3} (bottom panels),
both in presence (left panels) and in absence
(right panels) of [O/Fe] plateau.
A discontinuity is exhibited by the TGD at
$\phi=0.20$, where bulge formation is assumed
to start.   Though a G-dwarf problem seems to
exist for both the halo and bulge, the current
model provides a viable interpretation to the
occurrence of three extremum points, two
maxima and one minimum, in the EGD.

\section{Inhomogeneous, simple models}
\label{inmo}

As a viable alternative to homogeneous simple models,
let us take into consideration inhomogeneous simple
models already used in previous approaches (C00; C01).
In this view, the proto-halo/proto-bulge is conceived
as structured into a number of discrete, entirely
gaseous, identical regions, and a subsystem of
long-lived stars, remnants, and star formation
inhibiting gas,
which have been generated earlier. The evolution
occurs via a sequence of identical time steps. At the
beginning of a step, star formation stochastically takes
place in a subclass of ``active'' regions, as described
by simple homogeneous models, while the others remain
``quiescent''. At the end of a step, high-mass stars
have died whereas low-mass stars have survived upto today,
according to instantaneous recycling approximation.
In addition, the enriched gas left from active regions
is instantaneously mixed with the unenriched gas
within quiescent regions, to form a new set of identical
regions for the next step. For further details, see C00.

The gas oxygen abundance, averaged over the whole
system, still shows a monotonic increase with
time, and depends on the gas mass fraction,
according to Eq.\,(\ref{eq:Oyz}), provided the
fraction of active regions equals the probability of
a region to be active (expected evolution; for
further details, see C00).   Taking into consideration
also star formation inhibiting or enhancing gas,
the effective yield takes the expression
(C01):
\begin{lefteqnarray}
\label{eq:ypyq}
&& \hat{p}^\prime=\hat{p}^{\prime\prime}\frac
{(1-q)\mu_R^\prime\ln\mu_R^\prime}{(1-\mu_R^
\prime)q\ln q}~~;
\qquad\hat{p}^\prime\le\hat{p}^{\prime\prime}~~; \\
\label{eq:q}
&& q=1-\chi(1-\mu_R^\prime)~;\quad0\le q\le1~;
\end{lefteqnarray}
concerning an active region, $\hat{p}^{\prime
\prime}$ is the effective yield due to
star formation inhibiting or enhancing gas,
expressed by Eq.\,(\ref{eq:ypy}), 
and $\mu_R^\prime$ the gas mass fraction
at the end of a step; concerning the
system, $q$ may be thought of as an effective gas
mass fraction within a region at the end of a step,
i.e. the mean gas mass fraction averaged on both
active and quiescent regions, and $\chi$ is the
probability of a region being active at the beginning
of a step. For further details, see C00.

Though the TGD cannot be analytically expressed
in the framework of inhomogeneous simple models,
still it can be done in the special case of
expected evolution, with regard to the starting
point, $\psi(\phi_o)=\psi_o$, and the ending
point of the initial step, $\psi(\phi_o+\Delta
\phi_o)=\psi_1$, where $\Delta\phi_o$ is the net
oxygen abundance (normalized to the solar value)
increase in gas component at the end of the first
step.   The result is (C01):
\begin{lefteqnarray}
\label{eq:psiO}
&& \psi_o=\psi(\phi_o)=\log\left[\frac1N\left(\frac{\diff N}
{\diff\phi}\right)_{\phi_o}\right]=\log\left(\frac{\chi\mu_o}
{\mu_o-\mu_f}\frac{{\rm O}_\odot}{\hat{p}^{\prime\prime}}
\right)~~; \\
\label{eq:psi1}
&& \psi_1=\psi(\phi_o+\Delta\phi_o)=\log\left[\frac1N
\frac{\Delta N_o}{\Delta\phi_o}\right]\nonumber \\
&& \phantom{\psi_1}
=\log\left\{\frac1{\Delta
\phi}\left[1-\exp\left(-\frac{\Delta\phi}{\hat{p}}{\rm O}_\odot
\right)\right]\right\}+\log\frac{\chi\mu_o}{\mu_o-\mu_f}~~;
\end{lefteqnarray}
where $\Delta N_o$ is the number of stars belonging
to a selected spectral class, within the oxygen
abundance range, $\phi_o\le\phi\le\phi_o+\Delta\phi_o$,
and $\Delta\phi_o=\Delta\phi$ independent of the
step, in the case under discussion of expected
evolution.

An approximate expression of the TGD related to
the first step, $\psi_1$, with the terms up to
the second order retained, reads (C01):
\begin{equation}
\label{eq:ps1a}
\psi_1=\psi(\phi_o+\Delta\phi_o)=-\frac1{\ln10}\frac
{{\rm O}_\odot}{\hat{p}^{\prime\prime}}\frac{\Delta\phi}2+
\log\left(\frac{\chi\mu_o}{\mu_o-\mu_f}\frac{{\rm O}_\odot}
{\hat{p}^{\prime\prime}}\right)~~;\qquad\frac{{\rm O}_\odot}
{\hat{p}^{\prime\prime}}\frac{\Delta\phi}2\ll1~~;
\end{equation}
which is valid also in the general case, provided
$\Delta\phi$ is replaced by $\Delta\phi_o$ and the
probability, $\chi$, by the relative frequency,
$\nu_o=k_o/n_o$, being $k_o$ and $n_o$ the number
of active and all regions, respectively, with
regard to the first step.

The values of some parameters related to the expected
evolution, concerning cases H1, H2, B1, B2, are listed
in Tab.\,\ref{t:HBi}.
\begin{table}
\caption[pahd]{Values of parameters related
to the expected evolution of inhomogeneous
simple models, in connection
with two different cases for the halo, H1
and H2, and for the bulge, B1 and B2,
respectively.   The indices, 2.9 and 2.35,
denote values related to the power-law IMF
exponent, $p$, in computing the corresponding
quantities.   With regard to the parameter,
$\psi_1$, upper and lower values are
calculated by use of Eqs.\,(\ref{eq:psi1})
and (\ref{eq:ps1a}), respectively.   The
effective yield, $\hat{p}^\prime$, is
related to inhomogeneities in oxygen
abundance due to the presence of active
and quiescent regions, whereas oxygen is
uniformly distributed within active regions.
The lower part of the table is related to
models with star formation inhibiting or
enhancing gas.   Parameters not reported
therein have same value as in the upper part,
with the exception of $\hat{p}$, $\alpha$,
and $m_{mf}$, which are listed in Tab.\,\ref
{t:infi} together with other parameters not
appearing here.   The effective yield, $\hat
{p}^{\prime\prime}$, related to star formation
inhibiting or enhancing gas, is listed
as $\hat{p}$ in the upper part of the table.
The effective yield, $\hat{p}^\prime$, due to
the presence of both star formation inhibiting
or enhancing gas within active
regions, and star formation precluding gas
within quiescent regions, is listed with the
same notation in the upper part of the table.
The index, $R$, denotes a generic active
region.   The mean oxygen abundance
(normalized to the solar value) of stars
at the end of evolution is denoted as $\overline
{\phi}$.   Positive
and negative $\kappa$-$D$ values correspond to star
formation inhibiting and enhancing gas, respectively.
Mass fractions are normalized to
the initial mass.}
\label{t:HBi}
\begin{center}
\begin{tabular}{crrrr}
\multicolumn{1}{c}{} & \multicolumn{1}{c}{H1} &
\multicolumn{1}{c}{H2} & \multicolumn{1}{c}{B1} &
\multicolumn{1}{c}{B2} \\
\noalign{\smallskip}
\hline\noalign{\smallskip}
$\mu_R^\prime$ & 8.2201~E$-$4 & 4.0337~E$-$5 & 1.4139~E$-$1 & 8.5397~E$-$2 \\
$q$ & 4.4991~E$-$2 & 1.7298~E$-$3 & 2.4833~E$-$1 & 1.4548~E$-$1 \\
$\kappa$ & 9.5579~E$-$1 & 9.9831~E$-$1 & 8.7545~E$-$1 & 9.3431~E$-$1 \\
$\hat{p}/{\rm O}_\odot$ & 1.3363~E$-$1 & 9.2288~E$-$2 & 1.0857~E$-$0 &
8.1430~E$-$1 \\
$\hat{p}^\prime/{\rm O}_\odot$ & 5.3452~E$-$3 & 3.4180~E$-$3 &
7.6001~E$-$1 & 5.7001~E$-$1 \\
$\mu_f$ & 4.0973~E$-$6 & 8.9544~E$-$5 & 3.8030~E$-$3 & 4.4794~E$-$4 \\
$\alpha_{2.9}$ & 9.3914~E$-$1 & 9.5717~E$-$1 & 6.5510~E$-$1 & 7.1692~E$-$1 \\
$\alpha_{2.35}$ & 9.7832~E$-$1 & 9.8492~E$-$1 & 8.4739~E$-$1 & 8.8101~E$-$1 \\
$(\tilde{m}_{mf})_{2.9}$ & 6.7781~E$-$2 & 4.5917~E$-$2 & 4.6050~E$-$1 &
3.7068~E$-$1 \\
$(\tilde{m}_{mf})_{2.35}$ & 7.5651~E$-$5 & 2.6985~E$-$5 & 1.7263~E$-$2 &
8.7958~E$-$3 \\
$\Delta\phi$ & 1.6577~E$-$2 & 2.1738~E$-$2 & 1.0587~E$-$0 & 1.0988~E$-$0 \\
$\Delta\phi_R^\prime$ & 9.4927~E$-$1 & 9.3379~E$-$1 & 2.1239~E$-$0 &
2.0035~E$-$0 \\
$\psi_o$ & 8.5446~E$-$1 & 1.0341~E$-$0 & $-$9.1837~E$-$2 & 5.9898~E$-$2 \\
$\psi_1(\ref{eq:psi1})$ & 8.2781~E$-$1 & 9.8398~E$-$1 & $-$2.8650~E$-$1 &
$-$2.0066~E$-$1 \\
$\psi_1(\ref{eq:ps1a})$ & 8.2753~E$-$1 & 9.8297~E$-$1 & $-$3.0358~E$-$1 &
$-$2.3312~E$-$1 \\
\hline\noalign{\smallskip}
$\kappa$ & 4.5169~E$-$0 & 6.9883~E$-$0 & $-$3.2093~E$-$1 & $-$9.4658~E$-$2 \\
$s_R^\prime$ & 1.8111~E$-$1 & 1.2518~E$-$1 & 1.2645~E$-$0 & 1.0102~E$-$0 \\
$D_R^\prime$ & 8.1807~E$-$1 & 8.7478~E$-$1 & $-$4.0590~E$-$1 &
$-$9.5626~E$-$2 \\
$\overline{\phi}$ & 1.3385~E$-$1 & 9.3250~E$-$2 & 9.3597~E$-$1 &
8.2723~E$-$1 \\
$s_f$ & 1.8126~E$-$1 & 1.2518~E$-$1 & 1.4671~E$-$0 & 1.1041~E$-$0 \\
$D_f$ & 8.1874~E$-$1 & 8.7482~E$-$1 & $-$4.7094~E$-$1 & $-$1.0451~E$-$1 \\
\noalign{\smallskip}
\hline
\end{tabular}
\end{center}
\end{table}
The parameters appearing therein, which equal
their counterparts corresponding to homogeneous
models (Tabs.\,\ref{t:infi} and \ref{t:HB}),
must be connected with active regions, and
for this reason some corresponding output
parameters show different values.   For
further details, see Appendix A.   New
input parameters are: the effective yield,
$\hat{p}^\prime$, and the normalized oxygen
abundance increase at the end of a step
with regard to the whole system, $\Delta
\phi$, and to an active region, $\Delta
\phi_R^\prime$, respectively.

Concerning an active region, parameter
values at the end of a step are calculated
using a homogeneous simple model with
star formation inhibiting or enhancing gas.
Let $\mu_R^\prime$, $s_R^\prime$,
and $D_R^\prime$, be related to star
formation allowing gas, long-lived stars,
and star formation inhibiting or enhancing gas,
from or enhanced in forming stars,
respectively.   Taking star formation
inhibiting gas
and leaving other parameters unchanged,
makes no variation in the final gas
mass fraction, $\mu_f$, as it is
unrelevant if gas is frozen into
long-lived stars or inhibited from
forming stars.   Accordingly, the
final star plus star formation inhibiting
or enhancing gas mass fraction, $s_f+D_f=1-\mu_f$,
also remains unchanged.   On the other
hand, $s_R^\prime+D_R^\prime=1-\mu_R^
\prime$ within active regions at the
end of a step, where $s_R^\prime/D_R^
\prime=s_f/D_f$ necessarily holds.
The combination of the above relations
yields:
\begin{lefteqnarray}
\label{eq:ssR}
&& s_f=\frac{1-\mu_f}{1-\mu_R^\prime}s_R^\prime~~; \\
\label{eq:DDR}
&& D_f=\frac{1-\mu_f}{1-\mu_R^\prime}D_R^\prime~~;
\end{lefteqnarray}
which are listed in the lower part of
Tab.\,\ref{t:HBi}.

\begin{figure}[t]
\centerline{\psfig{file=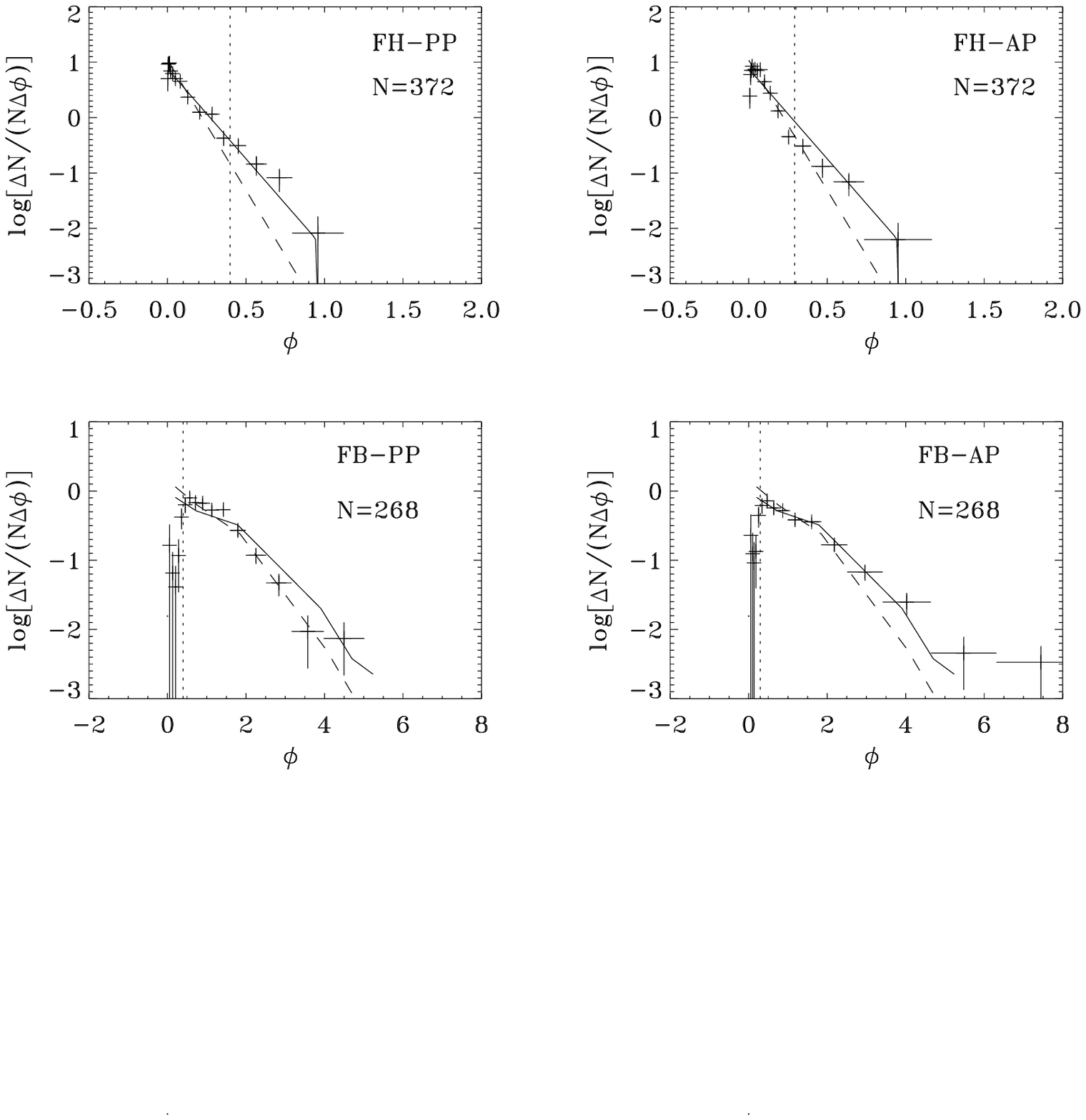,height=130mm,width=140mm}}
\caption[EGD]{Comparison between theoretical (TGD)
and empirical differential oxygen abundance
distribution (EGD) in halo (top panels) and bulge
(bottom panels) field stars,
both in presence (left panels) and in absence
(right panels) of [O/Fe] plateau, respectively.
Full curves correspond to inhomogeneous
models H1 (top panels) and B1 (bottom
panels), and dashed curves to H2 (top panels)
and B2 (bottom panels),
defined in Tab.\,\ref{t:HBi}.
Crosses represent the data and related
uncertainties, as in Fig.\,\ref{f:EGD4}.
The dotted vertical line marks the transition from
halo (OH, YH) to bulge/disk (BD) morphological
type in globular clusters, [Fe/H]=$-$0.8.}
\label{f:TGD6}
\end{figure}

The TGD related to cases H1, B1, (full
lines) and H2, B2, (dashed lines) is
compared in Fig.\,\ref{f:TGD6} to the
corresponding EGD with regard to
halo (top panels) and bulge
(bottom panels) field stars, both in presence
(left panels) and in absence (right
panels) of [O/Fe] plateau.

With regard to the halo, inhomogeneous simple
models with star formation inhibiting gas,
provide an acceptable fit using values of
input parameters listed in Tab.\,\ref{t:infi}
and $0.0033<\hat{p}^\prime/{\rm O}_\odot<0.0054$, in
presence and/or in absence of [O/Fe] plateau.

With regard to the bulge, inhomogeneous simple
models with star formation enhancing gas,
provide an acceptable fit using values of
input parameters listed in Tab.\,\ref{t:infi}
and $0.56<\hat{p}^\prime/{\rm O}_\odot<0.77$, in
presence and/or in absence of [O/Fe] plateau.

In dealing with active regions, the normalized
oxygen abundance at the end of evolution,
$\phi_f$, has to be replaced with the
normalized oxygen abundance at the end of
a step, $\phi_i+\Delta\phi_R^\prime$.

Further inspection of Fig.\,\ref{f:TGD6}
shows that, with respect to homogeneous
simple models plotted in Fig.\,\ref
{f:TGD4}, the fit is more or less
unchanged for the halo, and slightly
improved for the bulge.   In any case,
a G-dwarf problem still remains.

\begin{figure}[t]
\centerline{\psfig{file=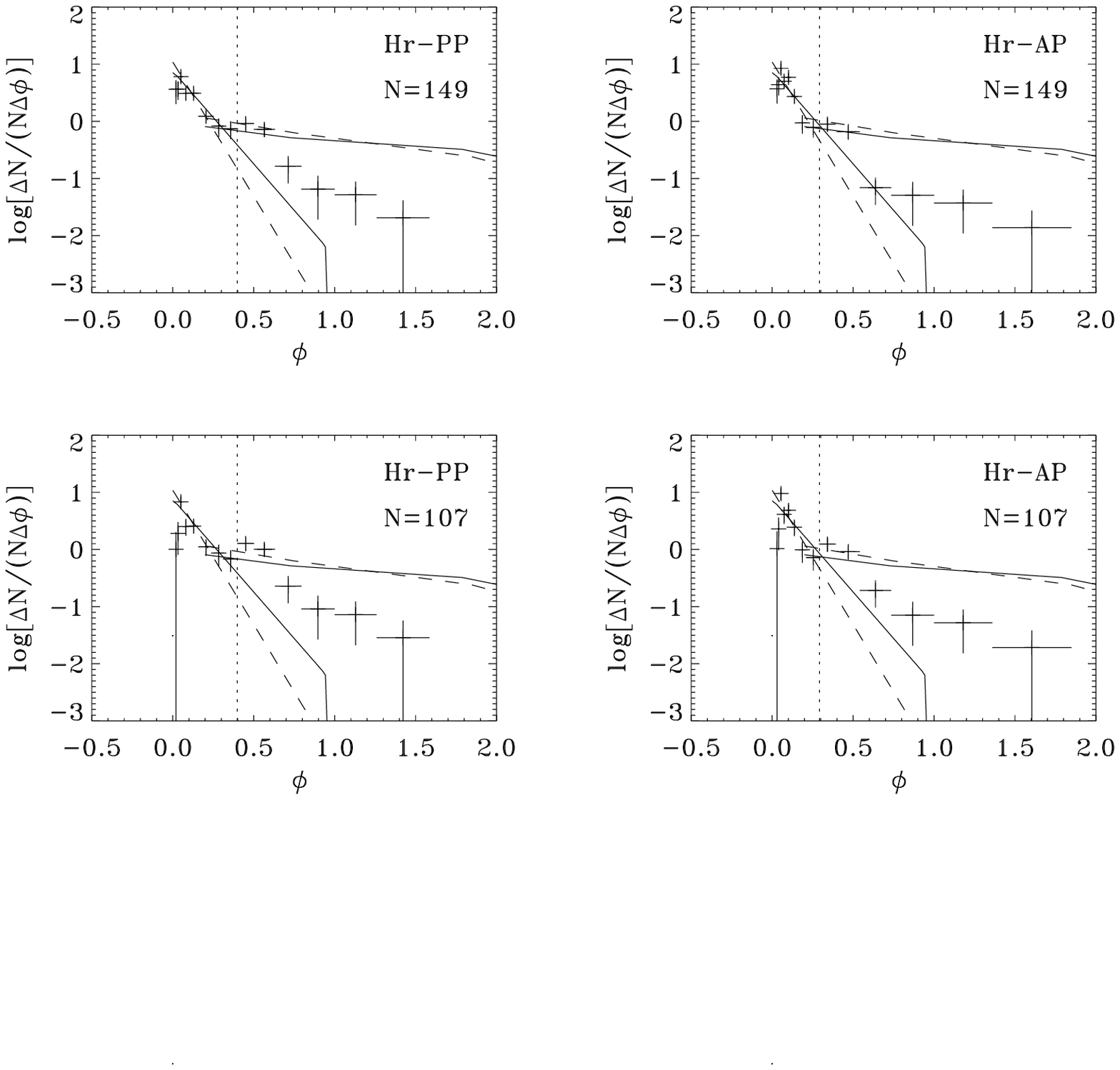,height=130mm,width=140mm}}
\caption[EGD]{Comparison between theoretical (TGD)
and empirical differential oxygen abundance
distribution (EGD) in all sample globular clusters
(top panels) and only old halo and bulge/disk
objects (bottom panels),
both in presence (left panels) and in absence
(right panels) of [O/Fe] plateau, respectively.
Full curves correspond to models H1 and B1,
and dashed curves to H2 and B2, where H curves
have larger slope with respect to B curves.
Crosses represent the data and related
uncertainties, as in Fig.\,\ref{f:EGD2}.
The dotted vertical line marks the transition from
halo (OH, YH) to bulge/disk (BD) morphological
type in globular clusters, [Fe/H]=$-$0.8.}
\label{f:TGD7}
\end{figure}

The TGD related to cases H1, B1, (full
lines) and H2, B2, (dashed lines) is
compared in Fig.\,\ref{f:TGD7} to the
corresponding EGD with regard to the
whole sample $(N=149)$ of globular
clusters considered above (top panels),
and a reduced sample $(N=107)$ of
old halo and bulge/disk objects
(bottom panels), both in presence
(left panels) and in absence (right
panels) of [O/Fe] plateau.   The
trend looks like its counterpart
exhibited by homogeneous simple
models, and similar considerations
can be made.

\begin{figure}[t]
\centerline{\psfig{file=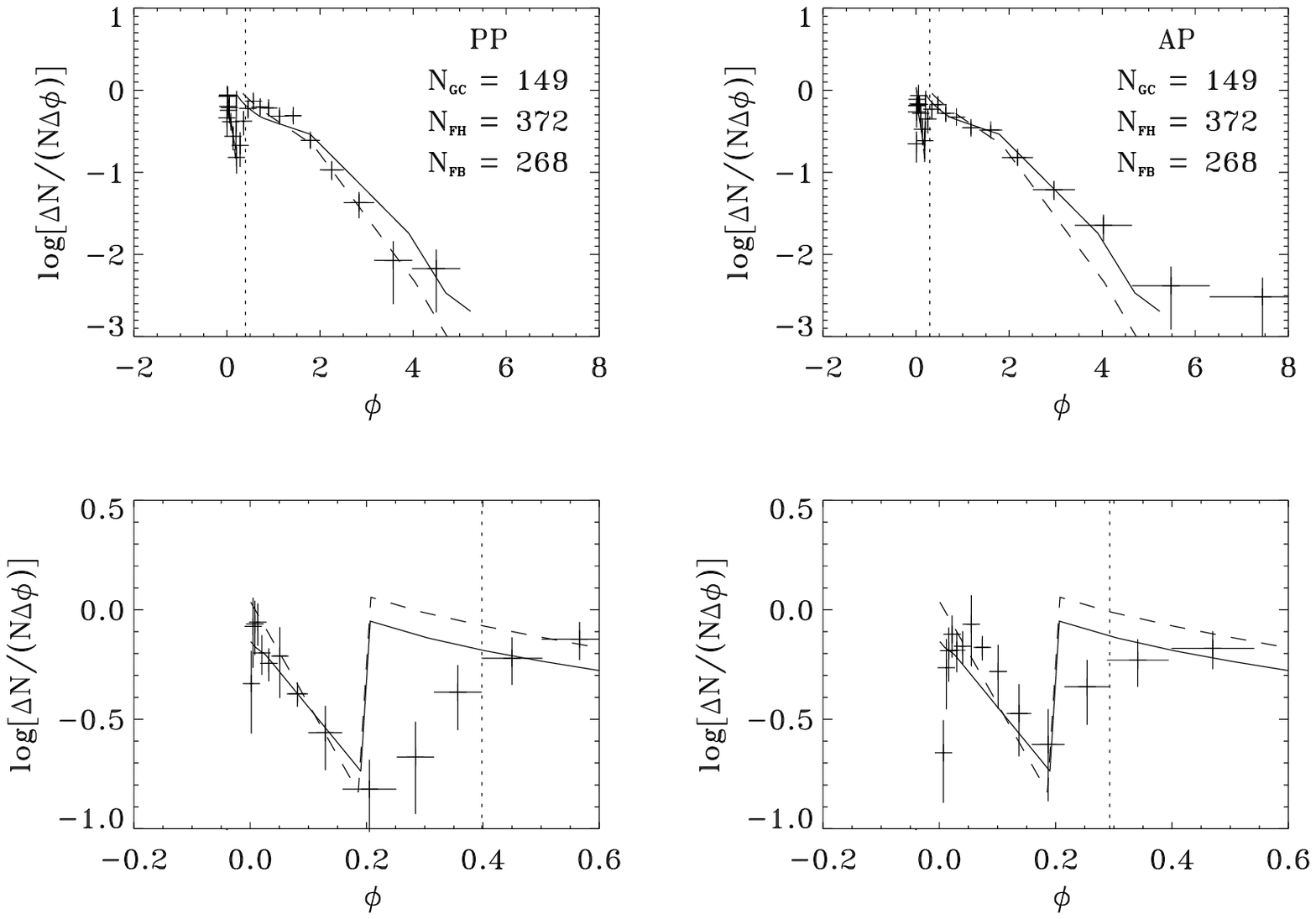,height=130mm,width=140mm}}
\caption[EGD]{Comparison between theoretical (TGD)
and empirical differential oxygen abundance
distribution (EGD) in the Galactic spheroid
(top panels) and zoomed for low normalized
oxygen abundances (bottom panels),
both in presence (left panels) and in absence
(right panels) of [O/Fe] plateau, respectively.
Full and dashed curves correspond to models H1,
B1; H2, B2; respectively, listed in Tabs.\,\ref
{t:infi}-\ref{t:HBi} and combined via Eq.\,(\ref
{seq:dpsi}).   Crosses represent the combination
of the data and related uncertainties via
Eqs.\,(\ref{seq:psis}), as in Fig.\,\ref{f:EGD3}
(bottom panels).
The dotted vertical line marks the transition from
halo (OH, YH) to bulge/disk (BD) morphological
type in globular clusters, [Fe/H]=$-$0.8.}
\label{f:TGD8}
\end{figure}

The TGD related to the Galactic spheroid has to be
numerically computed using Eq.\,(\ref{seq:dpsi})
together with H1, B1; H2, B2; models listed in 
Tab.\,\ref{t:HBi}, yielding cases S1, S2, respectively.
Fractional masses equal to $M_H/M=0.1$, $M_B/M=0.9$,
have been assumed to a good extent.   A comparison
with the related EGD (Fig.\,\ref{f:EGD3}, bottom
panels) is made in Fig.\,\ref{f:TGD8} both in
presence (left panels) and in absence (right panels)
of [O/Fe] plateau.   The trend looks like its
counterpart exhibited by homogeneous simple models,
and similar considerations can be made.

\begin{figure}[t]
\centerline{\psfig{file=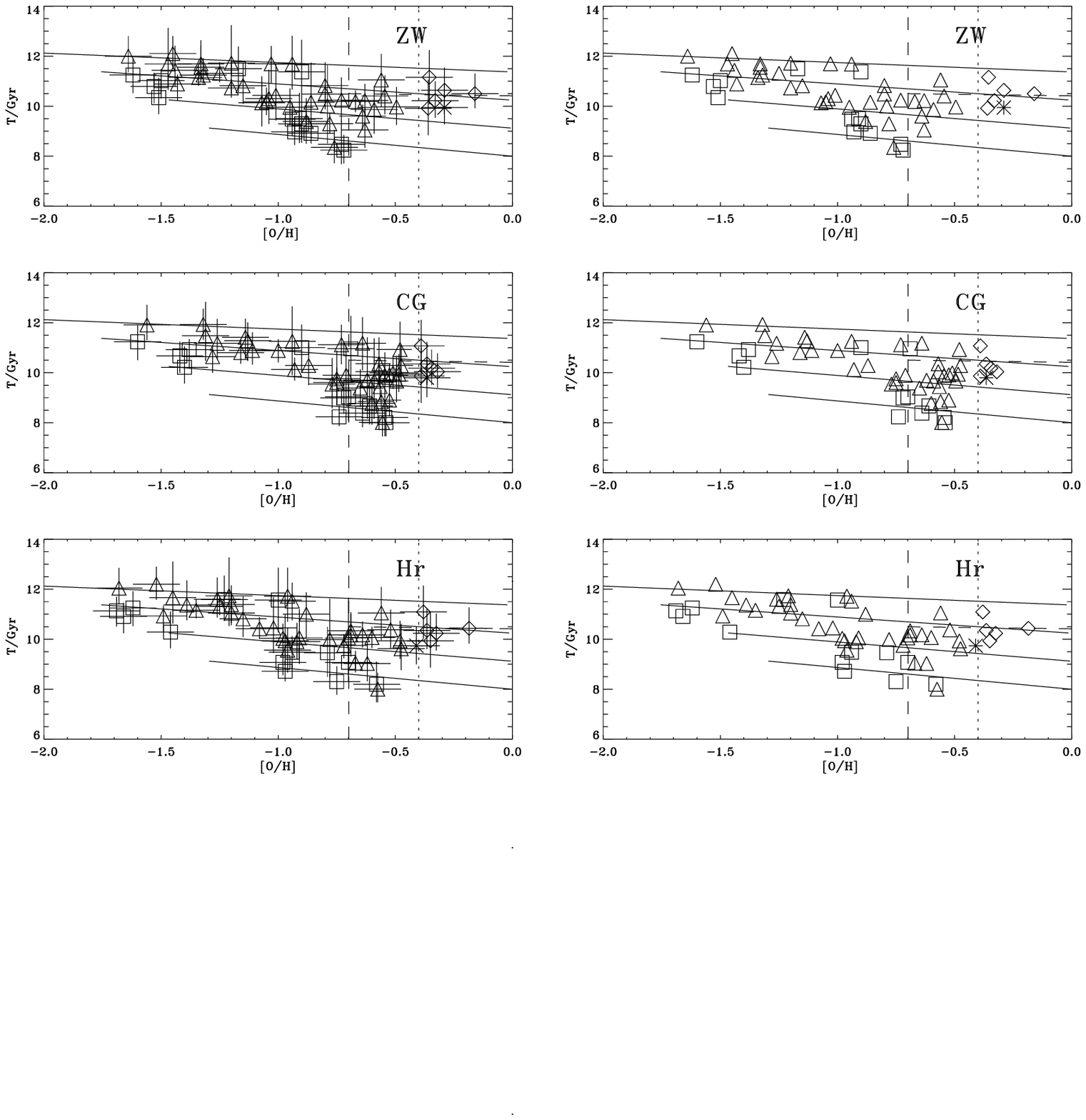,height=130mm,width=140mm}}
\caption[mamr]{Comparison between empirical (EAMR)
and theoretical age-metallicity relation
(TAMR) in presence of [O/Fe] plateau, according to
Eq.\,(\ref{eq:gra}).   The data come
from a sample of 55 globular clusters (De Angeli et
al. 2005) but expressed in terms of absolute ages (De
Angeli 2005).   Other captions as in Fig.\,\ref{f:MAMR}.
Full and dashed curves are related to models H1 and B1,
respectively.   Halo star formation begins at
([O/H], $T$/Gyr)$=$(-3, 12.5) and ends at (0, 8.0),
within four time steps, $\Delta T$/Gyr$=$1.125.
Bulge star formations begins at (-0.70, 10.5) and
ends at (0.74, 10.0), 
within four time steps, $\Delta T$/Gyr$=$0.125.
The last three steps are out of scale on the right,
and cannot be shown.
The dotted vertical line marks the transition from
halo (OH, YH) to bulge/disk (BD) morphological
type in globular clusters, [Fe/H]=$-$0.8.
The dashed vertical line marks the minimum in the
differential oxygen abundance distribution (EGD)
in the Galactic spheroid (Fig.\,\ref{f:TGD5}),
$\phi=0.20$ or $\log\phi=$[O/H]$=-$0.70.}
\label{f:OAMRP}
\end{figure}

\begin{figure}[t]
\centerline{\psfig{file=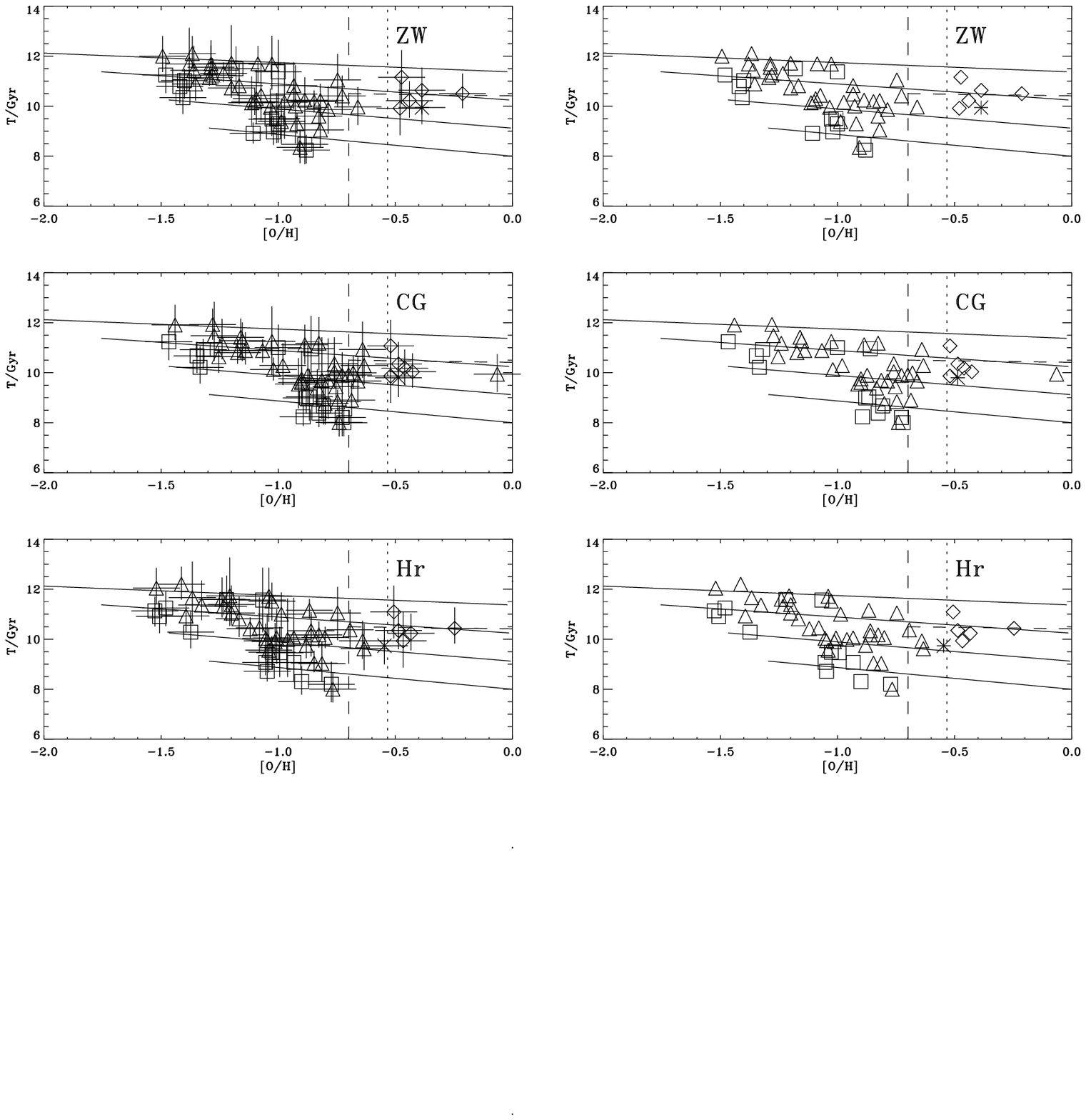,height=130mm,width=140mm}}
\caption[mamr]{Comparison between empirical (EAMR)
and theoretical age-metallicity relation
(TAMR) in absence of [O/Fe] plateau, according to
Eq.\,(\ref{eq:isa}).   The data come
from a sample of 55 globular clusters (De Angeli et
al. 2005) but expressed in terms of absolute ages (De
Angeli 2005).   Other captions as in Fig.\,\ref{f:MAMR}.
Full and dashed curves are related to models H1 and B1,
respectively.   Halo star formation begins at
([O/H], $T$/Gyr)$=(-$3, 12.5) and ends at (0, 8.0),
within four time steps, $\Delta T$/Gyr$=$1.125.
Bulge star formations begins at ($-$0.70, 10.5) and
ends at (0.74, 10.0), 
within four time steps, $\Delta T$/Gyr$=$0.125.
The last three steps are out of scale on the right,
and cannot be shown.
The dotted vertical line marks the transition from
halo (OH, YH) to bulge/disk (BD) morphological
type in globular clusters, [Fe/H]=$-$0.8.
The dashed vertical line marks the minimum in the
differential oxygen abundance distribution (EGD)
in the Galactic spheroid (Fig.\,\ref{f:TGD5}),
$\phi=0.20$ or $\log\phi=$[O/H]$=-$0.70.}
\label{f:OAMRA}
\end{figure}

The temporal behaviour of (allowing star formation)
gas oxygen
abundance (TAMR), normalized to the solar value,
$\log\phi=$[O/H], related to models H1 and B1, are
plotted in Figs.\,\ref{f:OAMRP} and \ref{f:OAMRA},
in presence and in absence of [O/Fe] plateau,
respectively.   Also plotted therein are the data
coming from a sample of 55 Galactic globular
clusters (De Angeli et al. 2005) but expressed
in terms of absolute ages (De Angeli 2005),
with same captions as in Fig.\,\ref{f:MAMR}.
The conversion from [Fe/H] to [O/H] has been
made using Eqs.\,(\ref{eq:gra}) and (\ref{eq:isa})
in presence and in absence of [O/Fe] plateau,
respectively.   It is worth recalling that
simple models of chemical evolution used in the
current paper are time independent, with regard
to the TGD.   Accordingly, the initial and the
final time, together with the time step in the
case of inhomogeneous models, can be selected
for best fitting the empirical age-metallicity
relation (EAMR).

With regard to the halo, initial and final
values, ([O/H], $T$/Gyr)$=(-$3, 12.5) and
(0, 8.0), respectively, have been chosen,
together with a time step, $\Delta T$/Gyr$=$
1.125, for a total of four.   The related TAMR
is represented by four full curves in
Figs.\,\ref{f:OAMRP} and \ref{f:OAMRA}.

With regard to the bulge, initial and final
values, ([O/H], $T$/Gyr)$=(-$0.70, 10.5) and
(0.74, 10.0), respectively, have been chosen,
together with a time step, $\Delta T$/Gyr$=$
0.125, for a total of four.   The related TAMR
is represented by a single full curve in
Figs.\,\ref{f:OAMRP} and \ref{f:OAMRA}, the
remaining three being out of scale on the
right.

The comparison between EAMR (related to
globular clusters) and TAMR (related to
both globular clusters and field stars)
shows that surviving (or formed) halo
globular clusters belong to all model
time steps, while surviving (or formed)
bulge/disk globular clusters belong to
the first step only and, in fact, are
coeval within the age uncertainties.

In the framework of inhomogeneous models,
star formation allowing gas has to be able of
generating both active and quiescent
regions.   With regard to models H and
B listed in Tab.\,\ref{t:HBi}, almost
all star formation allowing gas has been exhausted
at the end of evolution, leaving about
18-13\% of long-lived stars and 82-87\%
of star formation inhibiting gas in
the halo, and about 72-91\% of long-lived
stars from primeval gas and 28-9\% from
inflowing gas in the bulge, related to
the final mass.   The mean normalized
oxygen abundance within stars is about 
0.13-0.093 and 0.94-0.83 in the halo
and the bulge, respectively.   It is
worth of note that the Galactic spheroid (halo
and bulge) would be mostly luminous
for a IMF exponent, $p=2.9$, and mostly
dark for $p=2.35$, owing to different
lower stellar mass limits, by more than
an order of magnitude.   On the other
hand, the assumption of a universal IMF
implies the same yield, and lower stellar
mass limit, for the spheroid and the disk.

The comparison between homogeneous
(Tab.\,\ref{t:HB}) and inhomogeneous
\linebreak
(Tab.\,\ref{t:HBi}) simple models,
shows no substantial change in TGD.
It is worth emphasizing that the
only difference is in the physical
process related to the effective yield,
which is due to star formation inhibiting
or enhancing gas for
homogeneous models and, in addition,
to the presence of active and quiescent
regions for inhomogeneous models.
On the other hand, the related TAMR
cannot provide an acceptable fit in
the former alternative, but can do
in the latter, as shown in Figs.\,\ref
{f:OAMRP} and \ref{f:OAMRA}.

\section{Discussion} \label{disc}

According to the results of Sections 3 and 4,
halo and bulge EGD, taken separately, are
consistent with both homogeneous (Fig.\,\ref
{f:TGD4}) and inhomogeneous (Fig.\,\ref{f:TGD6})
simple models of chemical evolution.
The former alternative has been widely discussed
(e.g., Hartwick 1976; Caimmi 1981, 1982;
Ryan \& Norris 1991). The latter one allows
a direct comparison with its counterpart related
to the disk solar neighbourhood (C00), under the
assumption of a universal, power-law IMF.

As regards the disk solar neighbourhod, a lower 
stellar mass limit, exceeding the theoretical
Jeans stellar mass $(0.007\le\tilde{m}_J\le0.01)$,
occurs for a power-law exponent $p=-2.9$, which
is a fit to the Scalo (1986) IMF for $m\appgeq
{\rm m}_\odot$, concerning both mass distribution and
oxygen production (Wang \& Silk 1993). A less
steep Salpeter (1955) IMF, implying $p=-2.35$, 
is marginally consistent with the theoretical
Jeans stellar mass (Tab.\,\ref{t:infi}).   On
the other hand, the occurrence of stellar
wind would reduce oxygen nucleosynthesis by a factor of 
about 2.5 (e.g., Wang \& Silk 1993) which, in turn, would
raise the lower stellar mass limit (e.g., C01).

The more relevant parameters of inhomogeneous simple 
models related to halo, bulge, and disk solar neighbourhood,
are listed in Tab.\,\ref{t:phbd}.
\begin{table}
\caption[phbd]{Comparison between parameters
of inhomogeneous simple models, related to
halo and bulge (cases H1 and B1 of Tab.\,\ref
{t:HBi}) and disk solar neighbourhood (C01,
Tab.\,5).   An universal power-law initial
mass function (IMF) is assumed in all cases, which
leaves other parameters i.e. $\hat{p}$,
$\alpha$, and $m_{mf}$, unchanged.   Gas
and star mass fractions, $\mu_f$, $D_f$,
and $s_f$, are related to the initial mass
with regard to the bulge, where total mass
conservation is violated by gas inflow.}
\label{t:phbd}
\begin{center}
\begin{tabular}{lllll}
\multicolumn{1}{c|}{meaning}
&\multicolumn{1}{c|}{parameter}
&\multicolumn{2}{|c}{value} \\
\hline\noalign{\smallskip}
& & halo & bulge & disk \\
\noalign{\smallskip}
\hline\noalign{\smallskip}
probability & $\chi$ & $9.5579~10^{-1}$ & $\phantom{-}8.7545~10^{-1}$ &
$7.6765~10^{-2}$ \\
gas mass fraction & $\mu_R^\prime$ & $8.2201~10^{-4}$ &
$\phantom{-}1.4139~10^{-1}$ & $3.7148~10^{-1}$ \\
gas mass fraction & $q$ & $4.4991~10^{-2}$ & $\phantom{-}2.4833~10^{-1}$ &
$9.5175~10^{-1}$ \\
inhibition parameter & $\kappa$ & 4.5169 & $-3.2099~10^{-1}$ & 0.0000 \\
gas mass fraction & $\mu_f$ & $4.0973~10^{-6}$ &
$\phantom{-}3.8030~10^{-3}$ & $2.9047~10^{-1}$ \\
gas mass fraction & $D_f$ & $8.1874~10^{-1}$ & $-4.7094~10^{-1}$ & 0.0000 \\
star mass fraction & $s_f$ & $1.8126~10^{-1}$ & \phantom{$-$}1.4671 &
$7.0953~10^{-1}$ \\
normalized yield & $\hat{p}^{\prime\prime}/{\rm O}_\odot$ & $1.3363~10^{-1}$ &
$\phantom{-}7.6001~10^{-1}$ & $7.3722~10^{-1}$ \\
normalized yield & $\hat{p}^\prime/{\rm O}_\odot$ & $5.3452~10^{-3}$ &
\phantom{$-$}1.0857 & $4.4233~10^{-1}$ \\
\noalign{\smallskip}
\hline
\end{tabular}
\end{center}
\end{table}
The related values are taken from Tab.\,\ref
{t:HBi} (models H1 and B1) and C01 (Tab.\,5),
respectively.  An universal power-law IMF is
assumed, $\phi(\tilde{m})\propto\tilde{m}^-p$,
which makes no change in value of the physical
parameters, $\hat{p}$, $\alpha$, and $m_{mf}$.

With regard to active regions, the probability of
star formation is $\chi\appleq1$ for the halo,
$\chi<1$ for the bulge, and
$\chi\ll1$ for the disk. Accordingly, the gas
mass fraction left at
the end of a step, $\mu_R^\prime$, is close to
zero for the halo, about one seventh for the
bulge, and about one third for the
disk.  The effective yield, $\hat{p}^{\prime
\prime}$, is about six times larger in the
disk (where it coincides with the real yield)
and in the bulge (where star formation is
enhanced by inflowing gas with same composition
as in the pre-existing gas), than in the halo
(where gas is partially inhibited from forming
stars).

With regard to the whole system, a similar trend occurs.
The ratio of gas mass fraction at the end and at the
beginning of a step is $q\ll1$ for the halo, $q<1$ for
the bulge, and $q \appleq1$ for the disk. Accordingly,
the (allowing star formation)
gas mass fraction left at the end of evolution,
$\mu_f$, is close to zero for the halo and about
four thousandths for the bulge, and about
one third for the disk.  The effective yield,
$\hat{p}^\prime$, is about one half thousand
times larger in the bulge (where it is due to
both inhomogeneous star formation and gas inflow),
and about one thousand times larger in the disk
(where it is due to inhomogeneous star formation)
than in the halo (where it is due to both
inhomogeneous star formation and
inhibition from forming stars).
With regard to the total mass at the end of
evolution, it is left about 18\% of long-lived
stars (including remnants) and 84\% of gas
inhibited from forming stars in the halo;
about 72\% of long-lived stars from primeval
gas and 28\% from inflowed gas in the bulge;
about 71\% of long-lived stars and 29\% of
gas allowing star formation in the disk.

The TGD related to both homogeneous and
inhomogeneous simple models provides an
acceptable fit to the EGD related to the
Galactic spheroid (Figs.\,\ref{f:TGD5}
and \ref{f:TGD8}) and, in particular,
a non monotonic trend is reproduced.
While models assume that star formation
at the beginning of halo and bulge
evolution starts abruptly with constant
efficiency, the data seem to indicate
a somewhat gradual rate, with increasing
efficiency.   In other words, the
occurrence of some physical process
(not necessarily gas infall) seems
to inhibit halo and bulge star formation
in early times.   Accordingly, the
history of each Galactic subsystem
(halo, bulge, disk) could be conceived
as made of two distinct phases, namely
(i) assembling, where star formation
efficiency is gradually increasing,
and (ii) stabilization, where star
formation efficiency maintains constant.

At the end of halo evolution, the
fractional gas and star mass predicted
by the model, are:
\begin{lefteqnarray}
\label{eq:MHg}
&& \frac{M_{H\,{\rm gas}}}{M_{Ho}}=
\mu_f+D_f~~; \\
\label{eq:MHs}
&& \frac{M_{H\,{\rm stars}}}{M_{Ho}}=
s_f~~;
\end{lefteqnarray}
where $M_{Ho}$ is the initial halo mass.
The combination of Eqs.\,(\ref{eq:MHg})
and (\ref{eq:MHs}) yields:
\begin{equation}
\label{eq:MHgs}
M_{H\,{\rm gas}}=\frac{\mu_f+D_f}{s_f}
M_{H\,{\rm stars}}~~;
\end{equation}
where $M_{H\,{\rm stars}}$ is the current
halo mass.

At the end of bulge evolution, the fractional
gas plus stars and star mass due to gas inflow,
predicted by the model, are:
\begin{lefteqnarray}
\label{eq:MBg}
&& \frac{M_{B\,{\rm stars}}+M_{B\,{\rm gas}}}
{M_{Bo}}=\mu_f+s_f~~; \\
\label{eq:MBs}
&& \frac{M_{B\,{\rm stars}}^-}{M_{Bo}}=
-D_f~~;
\end{lefteqnarray}
where $M_{Bo}$ is the initial bilge mass and
$M_{B\,{\rm stars}}^-$ is the star mass due
to gas inflow, at the end of evolution.
The combination of Eqs.\,(\ref{eq:MBg})
and (\ref{eq:MBs}) yields:
\begin{equation}
\label{eq:MBgs}
M_{B\,{\rm stars}}^-=\frac{-D_f}{\mu_f+s_f}
(M_{B\,{\rm stars}}+M_{B\,{\rm gas}})~~;
\end{equation}
where $M_{B\,{\rm stars}}+M_{B\,{\rm gas}}$
is the current bulge mass.

With regard to models H1-2, B1-2, the values
of fractional masses, $\mu_f$, $s_f$, $D_f$,
are listed in Tab.\,\ref{t:HBi}, and taking
a current halo mass $M_H=M_{H\,{\rm stars}}=
10^9{\rm m}_\odot$ and a current bulge mass
$M_H=M_{B\,{\rm stars}}+M_{B\,{\rm gas}}=
10^{10}m_\odot$, it is found the following:
\begin{lefteqnarray}
\label{eq:MHm}
&& M_{H\,{\rm gas}}=4.5\range7.0~10^9{\rm m}_
\odot~~; \\
\label{eq:MBm}
&& M_{B\,{\rm stars}}^-=3.2\range0.95~10^
9{\rm m}_\odot~~;
\end{lefteqnarray}
which is in agreement with the idea, that
a fraction of the current bulge mass inflowed
from the halo.   The disk might have evolved
separately for two orders of reasons.

First, the empirical distribution of angular
momentum in halo, bulge, thick disk, and
thin disk stars seems to be consistent with
a decoupled dynamical evolution of the halo
and the thick disk i.e. dissipative halo-bulge
and thick disk-thin disk collapse (Wyse \&
Gilmore 1992; Ibata \& Gilmore 1995).
Accordingly, the chemical evolution of the
above mentioned subsystems could also have
been decoupled.

Second, the disk mass (gas + stars) is
estimated as $M_D=M_{D\,{\rm stars}}+M_{D\,
{\rm gas}}\approx(5.0+0.8)10^{10}{\rm m}_\odot=
5.8~10^{10}{\rm m}_\odot$ (e.g., Prantzos \&
Silk 1988), which cannot be provided
by outflowing halo gas for more than about
one tenth of the above value, according to
Eqs.\,(\ref{eq:MHm}) and and (\ref{eq:MBm}).

On the other hand, a continuous transition
seems to exist from an extended ($R\appgeq20$
kpc), pressure-supported halo, to an inner,
flattened ($R\appleq15$ kpc), rotation-supported
halo (Chiba \& Beers 2000). In addition, no
correlation appears between mean rotational
velocity and metal abundance for values below
[Fe/H]$\approx-1.7$ or $\phi\approx0.08$,
whereas exhibits a linear trend for larger values.
The mean rotational velocity exceeds the statistical
fluctuations related to metal-poor stars from [Fe/H]
$\approx-1.3$ or $\phi\approx0.2$ on, and the relative
abundance of thick disk stars is less than 1\% for
[Fe/H]$\approx-1.7$. For further details, see Chiba
\& Beers (2000).   It is worth noticing that the
range, $-1.7\appleq$[Fe/H]$\appleq-1.3$, with
regard to ZW metallicity scale, is related to a
larger age spread in globular clusters, as shown
in Fig.\,\ref{f:MAMR}.

The chemical abundance of thick disk stars suggests a
similar history to those of metal-rich ([Fe/H]$
\approx-1.3$) halo stars (Prochaska et al. 2000).
In addition, the thick
disk abundance patterns show excellent agreement
with the chemical abundances observed in metal-poor
bulge stars, suggesting the two populations were formed
from the same gas reservoir at a common epoch.
For further details, see Prochaska et al. (2000).

To include the disk in a qualitative discussion,
let us assume that a baryonic dark halo exists,
or an equivalent mass was lost during bulge and
disk formation (C01).   Accordingly, the following
relations hold:
\begin{leftsubeqnarray}
\slabel{eq:rM1a}
&& \frac{M_I}M=\frac{M_I}{M_{\rm vis}}\frac{M_{\rm vis}}
M~;\quad I=H,B,D ~; \\
\slabel{eq:rM1b}
&& M_{\rm vis}=M_H+M_B+M_D ~; \\
\slabel{eq:rM1c}
&& M=M_{\rm vis}+M_{\rm uns} ~;
\label{seq:rM1}
\end{leftsubeqnarray}
and, in addition:
\begin{equation}
\label{eq:rM2}
\frac{M_H+M_{\rm uns}}M=\frac{M_{\rm stars}}M~;
\quad\frac{M_B+M_D}M=\frac{M_{\rm gas}}M ~;
\end{equation}
where $M_{\rm vis}$ is the visible Galactic
mass, $M_{\rm uns}$ the unseen (baryonic)
mass, and $M_H+M_{\rm uns}$ is the total
halo mass, including the baryonic dark
subsystem or the lost mass.

Using the above quoted values for $M_H$,
$M_B$, $M_D$, it is found $M_H/M_{\rm vis}
=0.01449$, $M_B/M_{\rm vis}=0.14493$,
$M_D/M_{\rm vis}=0.84058$, and $M_{\rm vis} 
=6.9~10^{10}{\rm m}_\odot$.   On the other hand,
the values of fractional masses, $\mu_f$,
$s_f$, and $D_f$, predicted by the model,
must be related to the total mass, $M$.
Accordingly, the following relation holds:
\begin{equation}
\label{eq:MBD}
\frac{M_B+M_D}{M_{\rm vis}}=0.98551~~;
\end{equation}
via Eq.\,(\ref{eq:rM2}), and:
\begin{equation}
\label{eq:MHgM}
\frac{M_{H~{\rm gas}}}M=\mu_f+D_f~~;
\end{equation}
as predicted by the model via Eq.\,(\ref
{eq:MHg}), provided the initial halo mass
coincides with the initial Galactic mass.
If, in addition, bulge and (thick) disk
formation took place from the same gas
reservoir, as suggested by observations
(Prochaska et al. 2000), it may safely
be assumed that the proto-halo was the
common reservoir, which implies Eq.\,(\ref
{eq:rM2}).

The combination of Eqs.\,(\ref{eq:rM1a}),
(\ref{eq:rM2}), and (\ref{eq:MHgM}) yields:
\begin{equation}
\label{eq:MBDv}
\frac{M_B+M_D}{M_{\rm vis}}\frac{M_{\rm vis}}M=
\mu_f+D_f~~;
\end{equation}
where the mass fractions, $\mu_f$ and $D_f$,
are listed in Tab.\,\ref{t:HBi} with regard
to models H1 and H2.   Using Eqs.\,(\ref
{seq:rM1}) and (\ref{eq:MBD}), the following
mass ratios are determined:
\begin{leftsubeqnarray}
\slabel{eq:rM3a}
&& \frac{M_{vis}}M=0.83078\range0.88768~~; \\
\slabel{eq:rM3b}
&& \frac{M_{uns}}M=0.16922\range0.11232~~; \\
\slabel{eq:rM3c}
&& \frac{M_H}M=0.012040\range0.012865~~; \\
\slabel{eq:rM3d}
&& \frac{M_B}M=0.12040\range0.12865~~; \\
\slabel{eq:rM3e}
&& \frac{M_D}M=0.69834\range0.74618~~; \\
\slabel{eq:rM3f}
&& \frac{M_H}{M_{uns}}=0.071150-0.11454~~;
\label{seq:rM3}
\end{leftsubeqnarray}
which shows that the unseen (baryonic) halo,
or the total amount of gas lost, has to be
as massive as about the bulge.

In the light of the current model, it is apparent
that about 17\%\,-11\% of the total mass gave no
contribution to bulge and disk formation. If this
material is related to stars with a mass below a
treshold, $m_0$, then the mass ratio of long-lived
stars above and below the treshold must necessarily
equal the mass ratio, $M_H/M_{\rm uns}$.   The
further restriction to a power-law IMF, $\phi(\tilde
{m})\propto\tilde{m}^{-p}$, implies the validity of the
relation (C01):
\begin{equation}
\label{eq:MHu}
\frac{\tilde{m}_{mr}^{2-p}-\tilde{m}_{o}^{2-p}}
{\tilde{m}_{o}^{2-p}-\tilde{m}_{mf}^{2-p}}=
\frac{M_H}{M_{\rm uns}}~~;\qquad\tilde{m}=\frac m
{{\rm m}_\odot}~~;
\end{equation}
where $m_{mr}$ is the mass of the oldest halo
stars which are currently leaving the main
sequence.   For a model halo formation from
12.5 to 8.0 Gyr ago, $m_{mr}=0.9~{\rm m}_\odot$ is
expected to be consistent with the theory of
stellar evolution.   In fact, a value $m_{mr}
=0.87~{\rm m}_\odot$ is deduced by linear interpolation
from results related to metal-free stars
(Marigo et al. 2001), and the above value has
to be raised due to the observational lack of
halo stars with zero metallicity.

Taking $\tilde{m}_{mr}=0.9$, $\tilde{m}_{mf}$
from Tab.\,\ref{t:infi}, and using Eq.\,(\ref
{eq:rM3f}), the treshold mass, $\tilde{m}_o$,
may be deduced from Eq.\,(\ref{eq:MHu}).  The
result is:
\begin{leftsubeqnarray}
\slabel{eq:moa}
&& \tilde{m}_o=0.25048\range0.26178~~;\,\qquad~
p=2.9~~; \\
\slabel{eq:mob}
&& \tilde{m}_o=0.089514\range0.10027~~;\qquad
p=2.35~~;
\label{seq:mo}
\end{leftsubeqnarray}
accordingly, stars with lower mass (for fixed
power-law IMF exponent, $p$) either escaped
detection up today, or became unbounded to
the halo.

If, on the other hand, the unseen baryonic
halo is gaseous and unbound to the Galaxy,
the power-law IMF exponent related to $\tilde
{m}_{mf}=\tilde{m}_o$, leaving the remaining
input parameters unchanged, reads:
\begin{leftsubeqnarray}
\slabel{eq:pa}
&& p=2.8089\range2.8206~~; \\
\slabel{eq:pb}
&& p=2.6025\range2.6205~~;
\label{seq:p}
\end{leftsubeqnarray}
related to Eqs.\,(\ref{eq:moa}) and
(\ref{eq:mob}), respectively.
Accordingly, the values $p=2.9$ and
$p=2.35$ may be regarded as fiducial
limiting values.

An alternative explanation demands a different
IMF in different Galactic subsystems. A
minor change could only occur in the lower stellar
mass limit, owing to a larger Jeans stellar mass
(e.g., Larson 1998). Accordingly, a gas
amount prescribed to form stars in the range below
the Jeans stellar mass, $\tilde{m}_{mf}\le\tilde
{m}\le\tilde{m}_o$, would follow a different fate,
being lost from the system (e.g., Binney et al.
2001). The inhibition of star formation in the
mass range under discussion, could be due, in
addition, to less efficient cooling in metal-poor
proto-stars. In fact, the pre-main sequence life
time would be increased, to exceed the main
sequence life time of massive stars. Finally,
low-mass proto-stars would be destroyed by
supernovae and the related gas, which is expected
to be weakly bound to the proto-Galaxy, would be
lost during bulge and disk formation.

\section{Conclusion} \label{conc}

The empirical differential oxygen abundance
distribution (EGD) in the Galactic spheroid
has been deduced from three different samples
involving (i)
268 K-giant bulge stars (Sadler et al. 1996), and 
(ii) 149 globular clusters (Mackey \& van den Bergh
2005) for which the iron abundance distribution is
known, in addition to previous results (Caimmi 2001)
related to (iii) 372 solar neighbourhood halo
subdwarfs (Ryan \& Norris 1991).   To this aim,
two alternative [O/H]-[Fe/H] dependences have
been used, according to Eqs.\,(\ref{eq:gra}) and
(\ref{eq:isa}), respectively.   The data have
been fitted, to an acceptable extent, by both
homogeneous and inhomogeneous simple models of
chemical evolution.

Under the assumption of a universal initial
mass function (IMF) and same value of the
true yield as in the disk solar neighbourhood,
inhibition of halo star formation (implying
gas outflow) and enhanchement of bulge star
formation (implying gas inflow) have been
demanded for fitting the EGD.   On the contrary,
no such gas outflow or inflow was requested to
reproduce the EGD in the disk solar neighbourhood
(C00).   A power-law IMF has been considered,
$\phi(\tilde{m})\propto \tilde{m}^{-p}$, within
the range, $2.35\le p\le2.9$, and special effort
has been devoted to the limiting cases, $p=2.9$,
which is acceptably 
close to Scalo IMF for $m\appgeq {\rm m}_\odot$, and (ii)
an exponent $p=2.35$, which is the Salpeter IMF.
In any case, it has been inferred that a more
refined model involving an initially increasing
star formation efficiency (but not necessarily
implying gas infall) while assembling Galactic
subsystems, could provide a better agreement
with the data.

Homogeneous models have been recognized unable
in fitting the empirical age-metallicity relation
(EAMR) with regard to a homogeneous sample of
globular clusters De Angeli et al. 2005; De
Angeli 2005), which shows a non monotonic trend
characterized by large dispersion.   On the
other hand, inhomogeneous models have been
shown an acceptable fit, provided globular
cluster formation occurred through four different
steps in the halo and through a single step in
the bulge/disk, unless clusters of later generation
were disrupted.

With regard to gas outflow from the halo,
acceptable models made the following predictions.
If spheroid and disk component underwent distinct
evolutions, then a non negligible fraction of the
bulge mass (from about one third to about one
tenth) outflowed from the halo, for assumed $M_H=
10^9~{\rm m}_\odot$ and $M_B=10^{10}~{\rm m}_\odot$.   If, on
the other hand, spheroid and disk component
underwent a common evolution, then an unseen
baryonic halo (or equivalent amount of gas lost
by the Galaxy) has been shown to be needed, for
assumed $M_D=5.8 10^{10}~{\rm m}_\odot$.   The mass of
the unseen halo has been found to be of the same
order as the bulge mass.   In addition, the
treshold star mass below which the halo is not
detectable (or the stars are unbound to the
Galaxy) has been calculated as $m_o\approx0.25
~{\rm m}_\odot$ for IMF exponent $p=2.9$, and
$m_o\approx0.1{\rm m}_\odot$ for $p=2.35$;
conversely, $p\approx2.8$ for lower stellar
mass limit $m_{mf}=0.25~{\rm m}_\odot$, and
$p\approx2.6$ for $m_{mf}=0.1~{\rm m}_\odot$.

\section*{Acknowledgements}
We thank F. De Angeli for making available
values of absolute ages, errors, and additional
data, related to the quoted reference De Angeli
et al. (2005), and for stimulating e-mail
correspondence.   We also thank J. Fulbright
and M. Rich for making available a preprint
of the quoted reference Fulbright, McWilliam
\& Rich (2005) and, together with D. Terndrup,
for fruitful e-mail correspondence.

\appendix
\section*{Appendix}

\section{Correspondence between homogeneous and
inhomogeneous simple models}\label{a:corhi}

With regard to a generic system, let $\mu_o$,
$\mu_f$; $\phi_o$, $\phi_f$; be initial and
final values of fractional (allowing star
formation) gas mass and oxygen abundance
normalized to the solar value, respectively.
Let chemical evolution be described using
either homogeneous or inhomogeneous simple
models.  Finally, let the generalized yield,
$\hat{p}^{\prime\prime}$, related to the
system in the former alternative, coincide
with its counterpart related to an active
region in the latter alternative.

With regard to a generic active region
at the end of a step, Eq.\,(\ref{eq:Oyz})
reduces to (C01):
\begin{equation}
\label{eq:DpRp}
\Delta\phi_R^\prime=-\frac{\hat{p}^{\prime\prime}}
{{\rm O}_\odot}\ln\mu_R^\prime~~;
\end{equation}
where $\Delta\phi_R^\prime=\phi_{Rf}^\prime-
\phi_{Ro}^\prime$, $\mu_{Ro}^\prime=1$, and
$\mu_{Rf}^\prime=\mu_R^\prime$.   The
combination of Eqs.\,(\ref{eq:lgfi}), (\ref
{eq:Oyz}), and (\ref{eq:DpRp}), in the case
under discussion yields:
\begin{equation}
\label{eq:plmu}
\frac{\phi_f-\phi_o}{\Delta\phi_R^\prime}=
\frac{\ln(\mu_f/\mu_o)}{\ln\mu_R^\prime}~~;
\end{equation}
which allows the calculation of the fractional
gas mass ratio, $\mu_f/\mu_o$, predicted by
homogeneous simple models, in terms of
parameters related to inhomogeneous simple
models with equal values of initial and final
normalized oxygen abundance, $\phi_o$, and
$\phi_f$, respectively.

With regard to homogeneous simple models,
the mean normalized oxygen abundance in
long-lived stars at the end of evolution
is (C01, Appendix D):
\begin{equation}
\label{eq:fimo1}
\bar{\phi}=\bar{\phi}(\phi_f)=\phi_o+
(\phi_f-\phi_o)\frac{u(1-\ln u)-1}
{(1-u)\ln u}~~;\qquad u=\frac{\mu_f}{\mu_o}~~;
\end{equation}
and the combination of Eqs.~(\ref{eq:lgfi}),
(\ref{eq:Oyz}), and (\ref{eq:fimo1}) yields:
\begin{equation}
\label{eq:fimo2}
\bar{\phi}_f=\phi_0+\frac{\hat{p}^{\prime
\prime}}{{\rm O}_\odot}\left[1+\frac{u\ln u}{1-u}
\right]~~;\qquad u=\frac{\mu_f}{\mu_o}~~;
\end{equation}
where $u=\mu_f$ in the special case, $\mu_
o=1$.

With regard to inhomogeneous simple models,
the mean normalized oxygen abundance in
long-lived stars at the end of the
$\ell$-th step is (C01, Appendix D):
\begin{equation}
\label{eq:phimase}
\bar{\phi}_\ell=\frac{\sum_{i=0}^\ell\phi_i\nu_i\mu_i}
{\sum_{i=0}^\ell\nu_i\mu_i}+\frac{\mu_R^\prime(1-
\ln\mu_R^\prime)-1}{(1-\mu_R^\prime)\ln\mu_R^\prime}
\Delta\phi_R^\prime ~;
\end{equation}
where $\nu_i$ is the relative frequency of
active regions at the $i$-th step.

In the special case of expected evolution,
$\nu_i=\chi$, the first term on the right-hand
side of Eq.~(\ref{eq:phimase}) reduces to
(C01, Appendix D):
\begin{equation}
\label{eq:phimasa}
\frac{\sum_{i=0}^\ell\phi_i\nu_i\mu_i}
{\sum_{i=0}^\ell\nu_i\mu_i}=\phi_0-\frac
{\hat{p}^\prime}{{\rm O}_\odot}\frac{q\ln q}{1-q}
\left[1-\frac{(\ell+1)q^\ell(1-q)}{1-q^
{\ell+1}}\right]~~;
\end{equation}
where $\hat{p}^\prime$ is the generalized
yield related to the system, and $q$ may
be thought of as an effective gas mass
fraction within a region at the end of a
step i.e. the mean gas mass fraction
averaged on both active and quiescent
regions.   The combination of Eqs.~(\ref
{eq:ypyq}) and (\ref{eq:phimasa}) yields:
\begin{equation}
\label{eq:phimapp}
\frac{\sum_{i=0}^\ell\phi_i\nu_i\mu_i}
{\sum_{i=0}^\ell\nu_i\mu_i}=
\phi_0-\frac{\hat{p}^{\prime\prime}}{{\rm O}_\odot}
\frac{\mu_R^\prime\ln\mu_R^\prime}{1-\mu_R^
\prime}\left[1-\frac{(\ell+1)q^\ell(1-q)}
{1-q^{\ell+1}}\right]~~;
\end{equation}
in terms of the generalized yield, $\hat{p}^
{\prime\prime}$, related to an active region.
The substitution of Eqs.~(\ref{eq:DpRp}) and
(\ref{eq:phimapp}) into (\ref{eq:phimase})
produces:
\begin{equation}
\label{eq:phimi}
\bar{\phi}_\ell=\phi_0+\frac{\hat{p}^{\prime
\prime}}{{\rm O}_\odot}\left[1+\frac{\mu_R^\prime\ln
\mu_R^\prime}{1-\mu_R^\prime}\frac{(\ell+1)q^
\ell(1-q)}{1-q^{\ell+1}}\right]~~;
\end{equation}
at the end of the $\ell$-th step.

In the special case where the system reduces
to a single region and the evolution to a
single step, $\mu_f/\mu_o=\mu_R^\prime$,
$\ell=0$, Eq.~(\ref{eq:phimi}) reads:
\begin{equation}
\label{eq:phimio}
\bar{\phi}_\ell=\phi_0+\frac{\hat{p}^{\prime
\prime}}{{\rm O}_\odot}\left[1+\frac{\mu_R^\prime\ln
\mu_R^\prime}{1-\mu_R^\prime}\right]~~;
\end{equation}
which coincides with Eq.~(\ref{eq:fimo2}),
related to homogeneous simple models.

The above results may be reduced to a
single statement.
\begin{trivlist}
\item[\hspace\labelsep{\bf Theorem}] \sl
Given (i) an inhomogeneous simple model of
chemical evolution with assigned values
of initial normalized oxygen abundance,
$\phi_o$, generalized yield related to
an active region, $\hat{p}^{\prime\prime}$,
and gas mass fraction within an active
region at the end of a step, $\mu_R^\prime$;
and (ii) a homogeneous simple model of
chemical evolution with
initial normalized oxygen abundance,
$\phi_o$, generalized yield related to
the system, $\hat{p}^{\prime\prime}$,
and gas mass fraction at the end of
evolution, $\mu_f=\mu_R^\prime$; the
mean oxygen abundance in long-lived
stars at the end of the first step in
the former alternative, equals its
counterpart at the end of evolution in
the latter alternative.
\end{trivlist}

\end{document}